\documentclass{siamart0216}

\usepackage{lipsum}
\usepackage{amsfonts}
\usepackage{graphicx}
\usepackage{epstopdf}
\usepackage{algorithm,algpseudocode}
\usepackage{mwe,tikz}
\usetikzlibrary{positioning,calc,fadings,decorations.pathreplacing}
\usepackage{tikz-3dplot}
\usepackage{pgfplots}
\usepackage{caption}

\usepackage{algorithm}
\usepackage{algpseudocode}
\usepackage{pifont}

\usetikzlibrary{arrows, decorations.markings}
\usetikzlibrary{calc}
\usetikzlibrary{calc,shapes, positioning}
\usepackage{multicol}
\usepackage{tikz}
\usetikzlibrary{shapes,shadows,trees}
\ifpdf
  \DeclareGraphicsExtensions{.eps,.pdf,.png,.jpg}
\else
  \DeclareGraphicsExtensions{.eps}
\fi

\newcommand{\TheTitle}{A hybridizable discontinuous Galerkin method for two-phase flow in heterogeneous porous media} 
\newcommand{\TheAuthors}{M. S. Fabien, M. G. Knepley, and B. M. Rivi\`{e}re}

\headers{\TheTitle}{\TheAuthors}

\title{{\TheTitle}}

\author{
  Maurice S. Fabien\thanks{Department of Computational and Applied Mathematics, Rice University, Houston, TX.}
  \and
  Matthew G. Knepley  \thanks{Department of Computer Science and Engineering, State University of New York at Buffalo, Buffalo, NY.}
  \and
  B\'{e}atrice M. Rivi\`{e}re$^*$
}

\usepackage{amsopn}

\usepackage{bm}
\usepackage{booktabs}
\usepackage{xcolor,colortbl}
\usepackage{color}

\definecolor{Gray}{gray}{0.85}
\definecolor{LightCyan}{rgb}{0.88,1,1}

\newcolumntype{a}{>{\columncolor{Gray}}c}
\newcolumntype{b}{>{\columncolor{white}}c}

\ifpdf
\hypersetup{
  pdftitle={\TheTitle},
  pdfauthor={\TheAuthors}
}
\fi

\usepackage{subfigure}
 \usepackage[left=2cm,right=2cm,top=2cm,bottom=2cm]{geometry}
\begin{document}
\fontsize{10}{10}\selectfont
\maketitle

\begin{abstract}
We present a new method for simulating incompressible immiscible two-phase flow in porous media.  The semi-implicit method decouples  the wetting phase pressure and saturation equations.  The equations are discretized using a hybridizable discontinuous Galerkin (HDG) method.  The proposed method is of high order, conserves global/local mass balance, and the number of globally coupled degrees of freedom is significantly reduced compared to standard interior penalty discontinuous Galerkin methods.  Several numerical examples illustrate the accuracy and robustness of the method. These examples include verification of convergence rates by manufactured solutions, common 1D benchmarks and  realistic discontinuous permeability fields.
\end{abstract}


 

\section{Introduction}
Multiphase flows in porous media are fundamental processes in geophysics. For instance, they  characterize enhanced oil recovery \cite{peacemanbook}, hydrogeology \cite{helmig,bear2010modeling}, as well as CO2 sequestration in geological formations \cite{bielinski2007numerical}.   The equations that govern two-phase flow form a coupled system of nonlinear partial differential equations.  Numerous techniques have been proposed to resolve the nonlinearity inherent in the system, for instance, implicit-explicit (IMPES), semi-sequential, semi-implicit, and fully implicit methods \cite{chen2006computational}.
\\[12pt] 
The selection of spatial discretization is also a critical decision in the solution process.  Accuracy, local conservation, mass balance, and efficiency of implementation are all important features.  With respect to the wetting phase pressure equation, incorrect approximations to the phase velocity can cause oscillations and instability when used in the convection-dominated transport equation satisfied by the wetting phase saturation.  Compatible discretizations (as defined in \cite{dawson2004compatible}) for flow and transport maintains local and global mass conservation which provides stability and accuracy in the numerical methods.  
\\[12pt] 
In this work we discretize the pressure and saturation equations by the Hybridizable Discontinuous Galerkin (HDG) method.  The HDG method has several interesting properties. In particular, discrete analogs of global conservation for flow, and local conservation for transport are satisfied.  This postprocessing is available since the normal component of the numerical flux for the HDG method is single valued \cite{CockburnDGRS09}.  Classical discontinuous Galerkin (DG) methods construct the velocity from the pressure, which means that the velocity (and subsequent its $H(\text{div})$ postprocessing) converge sub-optimally \cite{BastianRiviere}.
\\[12pt]
An explosion of interest in discontinuous Galerkin methods has occurred since the 1990s.  DG methods have a number of attractive features, e.g. local mass conservation, $hp$--adaptation, their ability to handle nonconforming meshes, and the fact they are well-suited for parallelism.  One main disadvantage is that DG methods in general have more degrees of freedom than their continuous counterparts.   This difficulty is exacerbated by the fact that at each time step, multiple linear solvers may be required for complex porous media problems like two-phase flow (e.g. nonlinear solver, iterative coupling).  The hybridizable discontinuous Galerkin method addresses this concern \cite{CockburnDGRS09,CockburnGL09,CockburnDG08}.  A global system solely in terms of the approximate trace of the saturation variable can be obtained.  To do this, we prescribe a specific numerical flux for the approximate saturation variable, such that we can express it and the approximation to the saturation, in terms of an additional unknown defined on the skeleton of the mesh.  To ensure that the numerical trace is single valued, we require that the normal component of the numerical flux across the element boundaries is continuous. 
\\[12pt] 
For higher approximation orders, hybridization can significantly reduce the number of degrees of freedom \cite{samii2016parallel}.  In addition, the HDG method boasts optimal orders accuracy $k+1$ for \textit{all} approximate variables (including the velocity) if $k$ is the polynomial order. The method  possesses a local post-processing that can enhance the accuracy of the scalar variable (with an order of accuracy $k+2$), and retains favorable aspects of DG methods (e.g. local mass conservation, ability to handle unstructured meshes, etc.).  The piecewise constant case  is convergent for the Local Discontinuous Galerkin (LDG)-HDG (\cite{CockburnDG08}), as such, in the same framework, we can compare a finite volume type method to higher order schemes.  In addition, the piecewise constant case discretization is useful for multigrid \cite{fabien2}; as this system is much smaller and acts as a cheap coarse grid correction that smooths stubborn low frequencies in the multigrird process  The importance of hybridization cannot be overstated; at each time step one is required to perform multiple linear solves due to linearization (e.g. Newton's method) or iterative coupling (high order time stepping, velocity extrapolation).  This observation is even more important as one considers three dimensional multiphase flow problems.
\\[12pt]
DG methods have been applied to two-phase flow problems in porous media over the last ten years.  The most common techniques for resolving the nonlinearity present in the governing equations are IMPES (splitting pressure/saturation: implicit pressure, explicit saturation), time lagging (splitting pressure/saturation: implicit pressure, time lagged-implicit saturation), semi-implicit (splitting pressure/saturation:implicit pressure, implicit saturation), and fully implicit algorithms. 

IMPES methods are appealing computationally since the saturation equation is treated explicitly.  However, there are a number of drawbacks to this approach.  Such a method requires slope limiting, which may be complicated to formulate as no theory in 2D/3D exists.  Furthermore, although the saturation is treated explicitly, the time step may have to be taken prohibitively small in the presence of highly varying permeability.  DG methods using IMPES were pursued by authors in \cite{arbogast2013discontinuous,bastian2003higher,jamei2016novel}.  It is found that slope limiting is needed, as the saturation equation is of hyperbolic nature, and high order approximations lead to overshoot and undershoot that must be mitigated.  In \cite{klieber2006adaptive,ern2010discontinuous}, the authors  use a time lagging approach, which splits the pressure and saturation equations, but linearizes the saturation equation by time lagging its coefficients.  Slope limiting is needed in this approach as well, even though the time stepping is implicit (backward Euler).  A hybridized mixed finite element method coupled with DG  (IMPES) is considered in \cite{hoteit2008numerical}.  Slope limiting is required to stabilize the approximation of the hyperbolic saturation equation.  Another method is examined in \cite{hou2016adaptive}, with two different algorithms, namely, hybrid mixed finite element (HMFE)-DG (IMPES) and HMFE-DG (Picard-time lag).  Here the authors use iterative coupling when time lagging, and slope limiters are still required for stability.  All of the above IMPES/time lag methods use piecewise linear approximations for the saturation equation, as slope limiting reduces the accuracy to first order.
 \\[12pt]
 Fully implicit methods are the most computationally demanding, since they do not relax the nonlinearity in any way.  DG fully implicit approaches were explored in \cite{epshteyn2007fully,epshteyn2006solution,bastian2014fully}.  No slope limiters are required, but the resulting linear system that is to be solved at each time step is very large.  Polynomial orders up to four are considered in \cite{epshteyn2007fully} on very coarse meshes, and up to degree two are considered in \cite{bastian2014fully}.
 \\[12pt]
 To the best of our knowledge, our work presented in this paper is the first to explore high order HDG methods in a semi-implicit algorithm for multiphase flow (up to degree eight).  By splitting the pressure equation from the saturation equation, but treating both equations implicitly, we show that we do not need slope limiters.  Highly accurate approximations are available in a computationally efficient manner as hybridization significantly reduces the total number of degrees of freedom.  Primal DG methods are less accurate for the total velocity, as they need to reconstruct it by computing explicitely the gradient of the pressure approximation.  In contrast, the HDG method simultaneously obtains approximations for the wetting phase saturation, the gradient of wetting phase saturation, and the total velocity, that converge at the rate  $k+1$ in the $L^2$ norm.  Any issues with loss of mass balance do not give rise to instabilities or spurious oscillations as the HDG forms a pair of compatible discretizations.  Furthermore, local superconvergent postprocessing is available whenever an enhanced solution is required for the wetting phase pressure or saturation.

A brief outline of the paper is described. Section~\ref{sec:model} presents the model problem.  The HDG method is described in Section~\ref{sec:method} for both pressure and saturation equations. Numerical results are shown in Section~\ref{sec:numer}. Conclusions follow.

\section{Model problem}
\label{sec:model}
The equations that govern two-phase flow in a porous medium are given by a coupled system of nonlinear partial differential equations.  In this paper, we assume that the phases are incompressible and immiscible, which allows for the coupled system to be expressed in a pressure--velocity--saturation formulation.  That is, a system of partial differential equations for which the primary variables are the pressure and the saturation of the wetting phase and the total velocity:
\begin{align}
{\bf u}_t &= - \lambda_t  K \nabla   p_w-\lambda_o  K \nabla   p_c, & \quad \mbox{in} \quad\Omega\times (0,T),
\label{eq1}
\\
-\nabla \cdot {\bf u}_t &=0, & \quad \mbox{in} \quad\Omega\times (0,T),
\label{eq2}
\\
\frac{\partial (\phi s_w)}{\partial t} 
+ 
\nabla \cdot \bigg( \frac{\lambda_o \lambda_w}{\lambda_t}  K \nabla   p_c\bigg)
 &= 
 - \nabla \cdot \bigg( \frac{\lambda_w}{\lambda_t} {\bf u}_t \bigg), & \quad \mbox{in} \quad\Omega\times (0,T).
\label{eq3}
\end{align}
The domain $\Omega$ is the porous medium in consideration, and $(0,T)$ is the time interval.
The wetting (aqueous) phase pressure and saturation are denoted by $p_w$ and $s_w$, and similarly the non-wetting (oil) phase pressure and saturation are denoted by $p_o$ and $s_o$.  The capillary pressure, $p_c$, is the difference between the phase pressures. The total velocity is denoted by ${\bf u}_t$; it is the sum of the phase velocities.
The phase mobilities are denoted by $\lambda_w$ and $\lambda_o$. Capillary pressure and phase mobilities are nonlinear functions of the wetting phase saturation \cite{BrooksCorey}. The derivative of the capillary pressure is negative and we will rewrite $\nabla p_c$ as  $-|p_c'|\nabla s_w$ in the rest of the paper. 
The total mobility is denoted by $\lambda_t$. The sum of the phase saturations is equal to one. In summary, we have
\[
p_c = p_o - p_w, \quad \lambda_t = \lambda_w +\lambda_o, \quad s_w+s_o = 1.
\]
Permeability and porosity of the porous medium are denoted by $\phi$ and $K$ respectively. The permeability may vary over several orders in magnitude in heterogeneous media. 
\\[12pt]
Equations~(\ref{eq1})-(\ref{eq3}) are completed by initial condition for the saturation ($s_w = s_w^0$) and by various boundary conditions.  We decompose the boundary of the domain, $\partial \Omega$, into disjoint sets:
\begin{equation}
\partial \Omega = \Gamma_{p_\mathrm{D}} \cup \Gamma_{p_\mathrm{N}}
=
 \Gamma_{s_\mathrm{D}} \cup \Gamma_{s_\mathrm{N}}
,
~~~
\Gamma_{p_\mathrm{D}} \cap \Gamma_{p_\mathrm{N}}
=
\Gamma_{s_\mathrm{D}} \cap \Gamma_{s_\mathrm{N}}
=
\emptyset.
\end{equation}
We consider Dirichlet and Neumann boundary conditions for both wetting phase pressure and saturation, and total velocity. 
\begin{align}
p_w &= p_{\mathrm{D}},&\quad \mbox{on} \quad\Gamma_{p_\mathrm{D}}\times (0,T),
\\
{\bm u}_t \cdot {\bm n} &= 0,&\quad\mbox{on} \quad \Gamma_{p_\mathrm{N}}\times (0,T),
\\
s_w &= s_{\mathrm{D}},&\quad\mbox{on}\quad  \Gamma_{s_\mathrm{D}}\times (0,T),
\\
-K \frac{\lambda_w\lambda_o}{\lambda_t} p_c'\nabla s_w \cdot {\bm n} &= 0,&\quad\mbox{on}\quad  \Gamma_{s_\mathrm{N}}\times (0,T).
\end{align}
Our primary unknowns are the wetting phase pressure, saturation and total velocity. Other choices of primary unknowns can be made, for instance wetting phase pressure and non-wetting phase pressure, global pressure and wetting phase saturation \cite{bjornaraa2008comparing}.  We use the formulation given by equations~(\ref{eq1}), (\ref{eq2}), and (\ref{eq3}), as this formulation weakens the nonlinearity while retaining a middle ground in terms of computational efficiency, when compared to IMPES type methods (which require slope limiting) and fully implicit methods (more computationally expensive).

\section{The HDG method for two-phase}
\label{sec:method}
Let $\mathcal{E}_h$ be a regular mesh of the domain $\Omega$, made of triangular elements denoted by $E$. Let $\Gamma_h$ denote the set of interior and boundary edges.  We also denote
\[
\partial \mathcal{E}_h = \{ \partial E, \quad E\in\mathcal{E}_h\}.
\]
For any set $\mathcal{O}$, the short-hand notation $(\cdot,\cdot)_{\mathcal{O}}$ is used  for the $L^2$ inner-product over the elements of $\mathcal{O}$. For instance, if $\mathcal{O} = \Gamma_h$, we have
\[
(w,\mu)_{\Gamma_h}  = \sum_{e\in\Gamma_h} \int_e w \mu.
\]
\subsection{Pressure and velocity approximation}  
\label{sec:pressure}
Let $\mathbb{Q}_k(E)$ denote the space of polynomials of degree $k$ in each coordinate direction.  The HDG discrete spaces are introduced:
\begin{equation}
\begin{split}
W_h&= \{ w\in L^2(\Omega): w|_E \in \mathbb{Q}_{k}(E),~~~\forall E \in \mathcal{E}_h \},
\\
{\bm V}_h &= W_h \times W_h ,
\\
M_h &= \{ \mu \in L^2(\Gamma_h): \mu|_e \in \mathbb{Q}_k(e),~~~\forall e \in \Gamma_h \},
\\
M_h(0) &= \{ \mu \in M_h: \mu=0,~~~\text{on } \Gamma_{p_\mathrm{D}}\}
.
\end{split}
\end{equation}
The HDG discretization of equations~(\ref{eq1}) and~(\ref{eq2}) seeks $({\bm u}_{th}, p_{wh},\widehat{p}_{wh})\in {\bm V}_h \times W_h\times M_h(0)$ such that
\begin{align}
(\lambda^{-1}_t  K^{-1} {\bf u}_{th}, {\bf v})_{\mathcal{E}_h}
-
(   p_{wh}, \nabla\cdot {\bf v} )_{\mathcal{E}_h}
+
(\widehat{p}_{wh}, {\bf v}\cdot {\bm n})_{\partial\mathcal{E}_h}
&=
-
(\Pi_h p_{\mathrm{D}}, {\bf v}\cdot {\bm n})_{\Gamma_{p_\mathrm{D}}}
-
(\lambda_o \lambda_w  \lambda_t^{-1}    |p_c'(s_w)|\nabla s_w, {\bf v} )_{\mathcal{E}_h}
\label{eq:HDGdef1}
\\
-( {\bf u}_{th}, \nabla w)_{\mathcal{E}_h}
+
(\widehat{{\bf u} }_{th} \cdot {\bm n}, w)_{\partial\mathcal{E}_h}
 &= 0,
\label{eq:HDGdef2}  
\\
 ([\![ 
  \widehat{\bf u}_{th} \cdot {\bm n}
 ]\!]
  , \mu)_{\Gamma_h } 
 &=  0,
\label{eq:HDGdef3}
\end{align}
for all $({\bm v}, w,\mu)\in {\bm V}_h \times W_h\times M_h(0)$, where $\Pi_h$ denotes the $L^2$-projection onto $M_h$.  The numerical traces are given as follows:
\begin{align*}
\widetilde{p}_{wh}
&=
\begin{cases}
\widehat{p}_{wh},&\quad\text{on } 
\Gamma_h \backslash \Gamma_{p_\mathrm{D}}
\\
\Pi_h p_{\mathrm{D}},&\quad \text{on } 
\Gamma_{p_\mathrm{D}},
\end{cases}
\\
\widehat{\bf u}_{th}
&=
{\bf u}_{th}
+
\tau ( p_{wh} - \widetilde{p}_{wh}){\bm n},
\end{align*}
where $\tau$ is a piecewise constant stabilization defined on element boundaries.  In our work we set  \cite{Cockburn09}
\[
\tau|_e = K|_E\, \min_{{\bf x} \in e }{\lambda_t(s_{wh}({\bf x}))}, \quad \forall e\in \partial E,
\]
The HDG system written in matrix form can be expressed as
\begin{equation}
\def\arraystretch{1.5}
\begin{bmatrix}
A & -B^T & C^T
\\
B & D & E
\\
C & G & H
\end{bmatrix}
\begin{bmatrix}
U
\\
P
\\
\widehat{P}
\end{bmatrix}
=
\begin{bmatrix}
R_u
\\
R_p
\\
R_{\widehat{p}}
\end{bmatrix}
 \notag
,
\end{equation}
and isolating interior unknowns gives
\begin{equation}
\def\arraystretch{1.5}
\begin{bmatrix}
U
\\
P
\end{bmatrix}
=
\begin{bmatrix}
A & -B ^T
\\
B  & D 
\end{bmatrix}^{-1}
\Bigg(
\begin{bmatrix}
R_u
\\
R_p
\end{bmatrix}
-
\begin{bmatrix}
C^T
\\
E
\end{bmatrix}
\widehat{P}
\Bigg)
 \label{hdg_invert}
 .
\end{equation}
Due to the discontinuity of the approximations ${\bf u}_{th}$ and $p_{wh}$, the inverted matrix in equation~(\ref{hdg_invert}) can be performed in an element by element manner.  The equation that enforces continuity of  the normal component of the numerical trace of the total velocity is 
\begin{equation}
C U + GP + H \widehat{P} = R_{\widehat{p}}.
\end{equation}
We can condense the interior unknowns to obtain a globally coupled system only defined in terms of $\widehat{P}$, the wetting phase pressure on the mesh skeleton:
\[
\mathbb{H}\widehat{P} = \mathbb{F}, \quad
\mathbb{H}
 =
H - [C~G]
\begin{bmatrix}
A & -B ^T
\\
B  & D 
\end{bmatrix}^{-1}
\begin{bmatrix}
C^T
\\
E
\end{bmatrix},
\quad
\mathbb{F} =
R_{\widehat{p}}
-
[C~G]
\begin{bmatrix}
A & -B ^T
\\
B  & D 
\end{bmatrix}^{-1}
\begin{bmatrix}
R_u
\\
R_p
\end{bmatrix}.
\]
We note that the expressions for $\mathbb{H}$ and $\mathbb{F}$ can be obtained at the element level.  The HDG method \eqref{eq:HDGdef1}-\eqref{eq:HDGdef3}  has a number of appealing features, notably: 
\begin{itemize}
\item  the property that allows the element-by-element elimination of interior degrees of freedom, resulting in a significantly smaller fully coupled problem with its only unknowns on the mesh skeleton (this property is called static condensation), 

\item the approximations for ${\bm u}_{th}$, $p_{wh}$, and $\widehat{p}_{wh}$ all converge at the optimal rate of $k+1$,

\item a local element-by-element postprocessing for $p_{wh}$ results in a new approximation $p_{wh}^*$ that converges at the rate of $k+2$ \cite{CockburnDG08},

\item the numerical trace of ${\bm u}_{th}$ has its normal component continuous, which is a critical property for flows in heterogeneous media.
\end{itemize}

\subsection{Saturation approximation}
The transport equation~(\ref{eq3}) is typically convection dominated, with possibly degenerate parabolic nature. DG methods have shown promising results for miscible displacement and multiphase flows.  As such, a natural extension is to consider HDG methods because at each time step one may be required to perform multiple linear solves due to linearization (e.g. Newton's method) or iterative coupling (high order time stepping, velocity extrapolation).  In this section, we present an approximation for the wetting phase saturation by a hybridizable discontinuous Galerkin method.  We rewrite the transport equation~(\ref{eq3}) in first order form:
\begin{align}
{\bf q } - \nabla s_w &= 0,
\label{eq:sat1}
\\ 
\frac{\partial (\phi s_w)}{\partial t} 
 + \nabla \cdot ( {\bf F}_c(s_w) + {\bf F}_v(  {\bf q},s_w) ) &= 0,
\label{eq:sat2}
\end{align}
where ${\bf F}_c$ and ${\bf F}_v$ denote the convective and viscous terms:
\[
{\bf F}_c( {\bm u}_{t},s_w) = \frac{\lambda_w(s_w)}{\lambda_t(s_w)} {\bf u}_t,
 \quad 
 {\bf F}_v(  {\bf q},s_w) = -\frac{\lambda_o(s_w) \lambda_w(s_w)}{\lambda_t(s_w)} K |p_c'(s_w)| {\bf q}.  
\]
The continuous-in-time HDG discretization of \eqref{eq:sat1}-\eqref{eq:sat2} seeks $( {\bf q}_h, s_{wh},\widehat{s}_{wh})\in  W_h \times {\bm V}_h \times M_h$ such that
 \begin{align}
 \left({\bf q}_h,{\bf v}\right)_{\mathcal{E}_h} 
+
 \left(s_{wh},\nabla \cdot{\bf v}\right)_{\mathcal{E}_h} 
- 
\left(\widehat{s}_{wh},{\bf v}\cdot {\bm n}\right)_{\Gamma_h\setminus \partial\Omega} 
& = 
\left(\Pi_h s_{\mathrm{D}},{\bf v}\cdot {\bm n}\right)_{\Gamma_{s_\mathrm{D}}},
 \\
 \left(\frac{\partial (\phi s_{wh})}{\partial t} ,w\right)_{\mathcal{E}_h}
-
\left({\bf F}_c({\bm u}_{th},s_{wh}) + {\bf F}_v( {\bf q}_h,s_{wh} ) ,\nabla w\right)_{\mathcal{E}_h} 
+ 
\left( ({\widehat{\bf F}_c} (\widehat{\bm u}_{th},s_{wh},\widehat{s}_{wh}) + \widehat{\bf F}_v( {\bf q}_h,s_{wh},\widehat{s}_{wh}))\cdot {\bm n}  ,w\right)_{\partial\mathcal{E}_h} & = 0,
\\
\left(
(\widehat{\bf F}_c(\widehat{\bm u}_{th},s_{wh},\widehat{s}_{wh}) + \widehat{\bf F}_v( {\bf q}_h,s_{wh},\widehat{s}_{wh}))\cdot {\bm n}
,
 \mu
  \right)_{\Gamma_h\setminus\partial\Omega}
&= 0 ,
\label{eq:HDGdef4}
 \end{align}
for all $({\bf v}, w, \mu)  \in {\bm V}_h \times W_h\times M_h$.  The numerical convective and viscous fluxes are denoted $\widehat{\bf F}_c$ and $\widehat{\bf F}_v$ respectively. For $\widehat{\bf F}_c$, we use a Lax-Friedrich like numerical flux, in which we evaluate the analytical flux function with $\widehat{s}_{wh}$ and penalize the jump between $\widehat{s}_{wh}$ and $s_{wh}$.  In other words, we have
\begin{equation}
 \widehat{\bf F}_c( \widehat{\bm u}_{th},s_{wh},\widehat{s}_{wh})
= {\bf F}_c(\widehat{\bm u}_{th},\widehat{s}_{wh}) + \tau_c (s_{wh} - \widehat{s}_{wh}){\bm n},
 \notag
\end{equation}
 with $\tau_c>0$ as a stablization parameter \cite{Cockburn09,peraire2011embedded,fu2015analysis}.  In practice, given an edge $e$ on an element $E$, we choose
\[
\tau_c|_e = \max_{{\bf x} \in e }{ (| {\bf F}_c(  {\bm u}_{th}({\bf x}),s_{wh}({\bf x})) \cdot {\bm n}_e | , 0 )}.
\]
As we are using a LDG inspired HDG, we use the LDG flux (\cite{ArnoldBCM02,cockburn1998local}) for the diffusive numerical flux:
\begin{equation}
 \widehat{\bf F}_v( {\bf q},s_{wh},\widehat{s}_{wh} )
= 
{\bf F}_v( {\bf q}, \widehat{s}_{wh} ) 
+ 
\tau_v (s_{wh} - \widehat{s}_{wh}) {\bm n},
 \notag
\end{equation}
 with $\tau_v>0$ as a stablization parameter \cite{Cockburn09,peraire2011embedded,fu2015analysis}.  For an edge $e$ on an element $E$, we choose 
\[
\tau_v|_e = \min_{{\bf x} \in e }{  ( {\bf F}_v({\bm q}_h,s_{wh}({\bf x})), 1)  }.
\]
A further discussion on the selection and analysis of stabilization parameters can be found in~\cite{bui2015godunov}.  We mention briefly that the HDG stabilization factor does not depend on the polynomial order or mesh size, unlike other DG methods.
\\[12pt]
To discretize the temporal dimension we employ first order implicit Euler. We remark that high order BDF and implicit Runge Kutta methods have been used successfully for HDG methods \cite{jaust2014temporally,nguyen2012hybridizable}.
The above system describes the HDG discretization.  It is nonlinear, so at each time step we linearize it using Newton's method.  The first time step requires a sufficiently accurate initial guess.  To obtain this initial guess, we take one step of an Anderson accelerated Picard iteration \cite{anderson1965iterative}.  For all other time steps the initial guess is taken as the wetting phase saturation from the previous time step.
 \\[12pt]
 A key feature of the HDG method is its ability to utilize static condensation.  At a given time step our system in matrix form can be written as
\begin{equation}
\def\arraystretch{1.5}
\begin{bmatrix}
A_{qq} & A_{qs} & B_{q}
\\
A_{sq}  & A_{ss} & B_{s}
\\
C_{q}  & C_s & D
\end{bmatrix}
\begin{bmatrix}
Q
\\
S
\\
\widehat{S}
\end{bmatrix}
=
\begin{bmatrix}
R_q
\\
R_s
\\
R_{\widehat{s}}
\end{bmatrix}
,
\notag
~~~
A = 
\begin{bmatrix}
A_{qq} & A_{qs}  
\\
A_{sq}  & A_{ss}  
\end{bmatrix},
~~~
B =[B_q,B_s]^T,
~~~
C =[C_q,C_s]
,
\end{equation}
and condensing out the interior degrees of freedom $Q$ and $S$ gives 
\begin{equation}
H \widehat{S} = R, ~~~
H = D - C A^{-1} B,~~~ R = R_{\widehat{s}} - CA^{-1}  
\begin{bmatrix}
R_q
\\
R_s
\end{bmatrix}
 .
\notag
\end{equation}
Thus, one can assemble a globally coupled system that only depends on the trace unknowns $\widehat{s}_{wh}$.  We note that this assembly can be performed in an element by element manner due to the local nature of HDG method.  After the wetting phase saturation is solved for on the mesh skeleton, one may recover the volume space approximations through a element by element procedure.  Indeed, the Shur complement yields
\begin{equation}
\def\arraystretch{1.5}
\begin{bmatrix}
Q
\\
S
\end{bmatrix}
=
\begin{bmatrix}
A_{qq} & A_{qs}
\\
A_{sq} & A_{ss}
\end{bmatrix}^{-1}
\bigg(
\begin{bmatrix}
R_q
\\
R_s
\end{bmatrix}
-
B \widehat{S}
\bigg).
\notag
\end{equation}
It should be realized that $\widehat{S}$, $S$ and $Q$ are actually Newton increments, and not necessarily the desired solution, so one would have to introduce the following variables:
\begin{equation}
\begin{split}
{\widehat{S}}^{m+1}
&=
 \widehat{S}^{m}+ \widehat{S},
 \\
S^{m+1}
&=
 S^{m}+ S
 \\
Q^{m+1}
&=
 Q^{m}+ Q,
 \end{split}
\notag
\end{equation}
where superscripts on $\widehat{S}$, $S$ and $Q$ denote the $m$th Newton iteration.  To further accelerate convergence of Newton's method, a sufficient initial guess is required.  Anderson acceleration is one strategy that can be used.  It is also feasible to consider using a damped Newton iteration (for a HDG specific example see \cite{woopen2014hybridized}) in conjunction with Anderson acceleration to improve the convergence.  For our numerical experiments, one step of an Anderson accelerated Picard iteration allowed Newton's method to converge quadratically.
\\[12pt]
The HDG method for the saturation has the same properties as mentioned in subsection~\ref{sec:pressure} for the pressure-velocity system.  In particular, we reiterate that the approximations for ${\nabla s}_{wh}$, $s_{wh}$, and $\widehat{s}_{wh}$ all converge at the optimal rate of $k+1$.  
\subsection{Postprocessing the saturation variable}
\label{sec:postproc}
The postprocessing procedure used in this work is explained here, and is inspired by the one established in~\cite{nguyen2009implicit}.  The element-by-element postprocessing of the saturation $s_{wh} $ is denoted by $s_{wh}^* $; and results in a new piecewise discontinuous polynomial approximation of degree $k+1$ such that
\begin{align}
( \nabla s_{wh}^*, \nabla w)_{ E } &= (  {\bm q}_h, \nabla w)_{ E }, \quad \forall w\in \mathbb{Q}_{k+1}(E),
\\
(  s_{wh}^*,1)_{ E } &= (s_{wh},1)_{ E },
\label{eq:post_proc}
\end{align}
for all $E \in \mathcal{E}_h$, where we assume that ${\bm q}_h$ and $s_{wh}$ are known.  Numerical evidence (see section~\ref{sec:ex_Manu}) shows that this postprocessing converges for $k>0$, at the rate of $k+2$ in the $L^2$-norm, and $k+1$ in $H^1$-norm.  We mention that various postprocessings exist, some of which have their resulting approximation not satisfying the original PDE in any sense \cite{cockburn2009hybridizable_sacco,cockburn2009superconvergent,stenberg1991postprocessing,CockburnDG08}.  Moreover, to guarantee superconvergence on Cartesian meshes, it is necessary to use a slightly larger finite element space than the standard tensor product space \cite{cockburn2012conditions}.  The postprocessing does not need to occur at every time step, it can be activated at whenever an enhanced solution is desired.  Superconvergence of this postprocessing (steady-state or otherwise) is reliant on both $s_{wh}$ and $\nabla s_{wh}$ converging optimally at the rate of $k+1$ in the $L^2$-norm~\cite{nguyen2009implicit}.  The system in equation~\eqref{eq:post_proc} is element local, and as such, is completely data parallel.  Furthermore, it is cheaper to compute than a fully coupled linear system \cite{fabien2}.
\subsection{Semi-implict HDG algorithm}
The semi-implicit HDG two-phase flow algorithm can be found in Algorithm~\ref{alg:HDG}.  Let $t_n$ denote the time at the $n$th time step, so that $t_{n+1}=t_n+\Delta t$, where $\Delta t$ is a given time spacing.  The variable $n_{\text{steps}}$ is the total number of time steps required.  In addition, a superscript of $n$ means that the variable is evaluated at $t_n$, e.g., $  s_{wh}({\bf x},t_n) :=   s_{wh}^n$.  Similarly, for the Newton steps, a superscript of $m$ denotes the $m$th Newton iteration.
 \begin{center}
    \begin{minipage}{.6\linewidth} 
\begin{algorithm}[H]
\centering
\caption{Semi-implicit HDG two-phase flow.}
\label{alg:HDG}
\begin{algorithmic}[1]
\For{ $n=0$ to ($n_{\text{steps}}-1)$ }

\State \parbox[t]{\dimexpr\linewidth-\algorithmicindent}{Given ${\bm q}_{h}^n$ and $s_{wh}^n$, solve equations~\eqref{eq:HDGdef1},~\eqref{eq:HDGdef2}, and~\eqref{eq:HDGdef3} for ${\bm u}_{th}^n$, $p_{wh}^n$, and $\widehat{p}_{wh}^n$.}

\State \parbox[t]{\dimexpr\linewidth-\algorithmicindent}{
Set $({\bm q}_{h}^m,s_{wh}^m ,\widehat{s}_{wh}^m ) = ({\bm q}_{h}^n,s_{wh}^n ,\widehat{s}_{wh}^n )$,
}
\State Set $r = 1$,
    \While{ $r\ge $ tol}
    
\State \parbox[t]{\dimexpr\linewidth-\algorithmicindent}{
Using $({\bm q}_{h}^m, s_{wh}^m, \widehat{s}_{wh}^m)$ and ${\bm u}_{th}^n$, solve equation~\eqref{eq:HDGdef4} for $({\bm q}_{h}^{m+1},s_{wh}^{m+1},\widehat{s}_{wh}^{m+1}).$     
}   

\State \parbox[t]{\dimexpr\linewidth-\algorithmicindent}{
 Set $r =  \max\{ 
 \| {\bm q}_{h}^{m+1} -{\bm q}_{h}^m \|_{L^2(\Omega)},
  \| s_{wh}^{m+1} -s_{wh}^m \|_{L^2(\Omega)} ,
  \newline
  \| \widehat{s}_{wh}^{m+1} -\widehat{s}_{wh}^m \|_{L^2(\Omega)} 
  \}$,
 }  
 \State \parbox[t]{\dimexpr\linewidth-\algorithmicindent}{
 Set $  ({\bm q}_{h}^{m},s_{wh}^{m},\widehat{s}_{wh}^{m}) = ({\bm q}_{h}^{m+1},s_{wh}^{m+1},\widehat{s}_{wh}^{m+1}),$
 }
     \EndWhile
 \State Set $  ({\bm q}_{h}^{n},s_{wh}^{n},\widehat{s}_{wh}^{n}) = ({\bm q}_{h}^{m+1},s_{wh}^{m+1},\widehat{s}_{wh}^{m+1}),$
\EndFor
\end{algorithmic}
\end{algorithm}     
 \end{minipage}
 \end{center}
 
\section{Numerical experiments} 
\label{sec:numer}
In this section, several numerical experiments are examined.  We validate our method on two benchmark problems, the McWhorter and Buckley-Leverett problems. The McWhorter problem models counter-current two-phase flow where capillary forces are present \cite{mcwhorter1990exact}. The Buckley-Leverett problem is a well-known example of 1D hyperbolic transport \cite{buckley1942mechanism}. Both the McWhorter and Buckley-Leverett problems have analytic or semi-analytic solutions.  We verify that the correct convergence rates are obtained in 2D by using the method of manufactured solutions. Finally we test our approach on heterogeneous porous media, where no analytic solution is known.

\subsection{McWhorter problem}
 \label{sec:ex_Mc}
\begin{figure}[t!]
\centering
\includegraphics[trim = 10mm 85mm 20mm 85mm, clip, scale = 0.45]{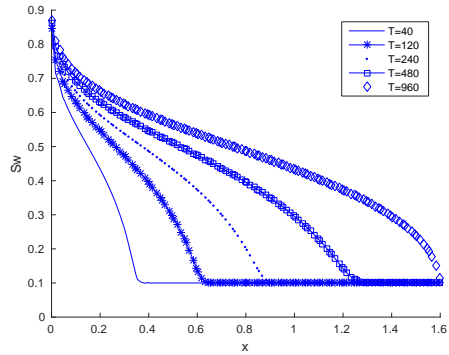} 
\caption{Saturation profiles at different times for the McWhorter problem.  Piecewise octic polynomials are used on a coarse mesh with 32 elements.}
\label{fig:mcWhorter_profile}
\end{figure}

\begin{figure}[t!]
\centering
    \captionsetup{justification=centering}
    \subfigure[\small{$k=0$}]{
        \includegraphics[trim = 40mm 80mm 40mm 90mm, clip, scale = 0.4]{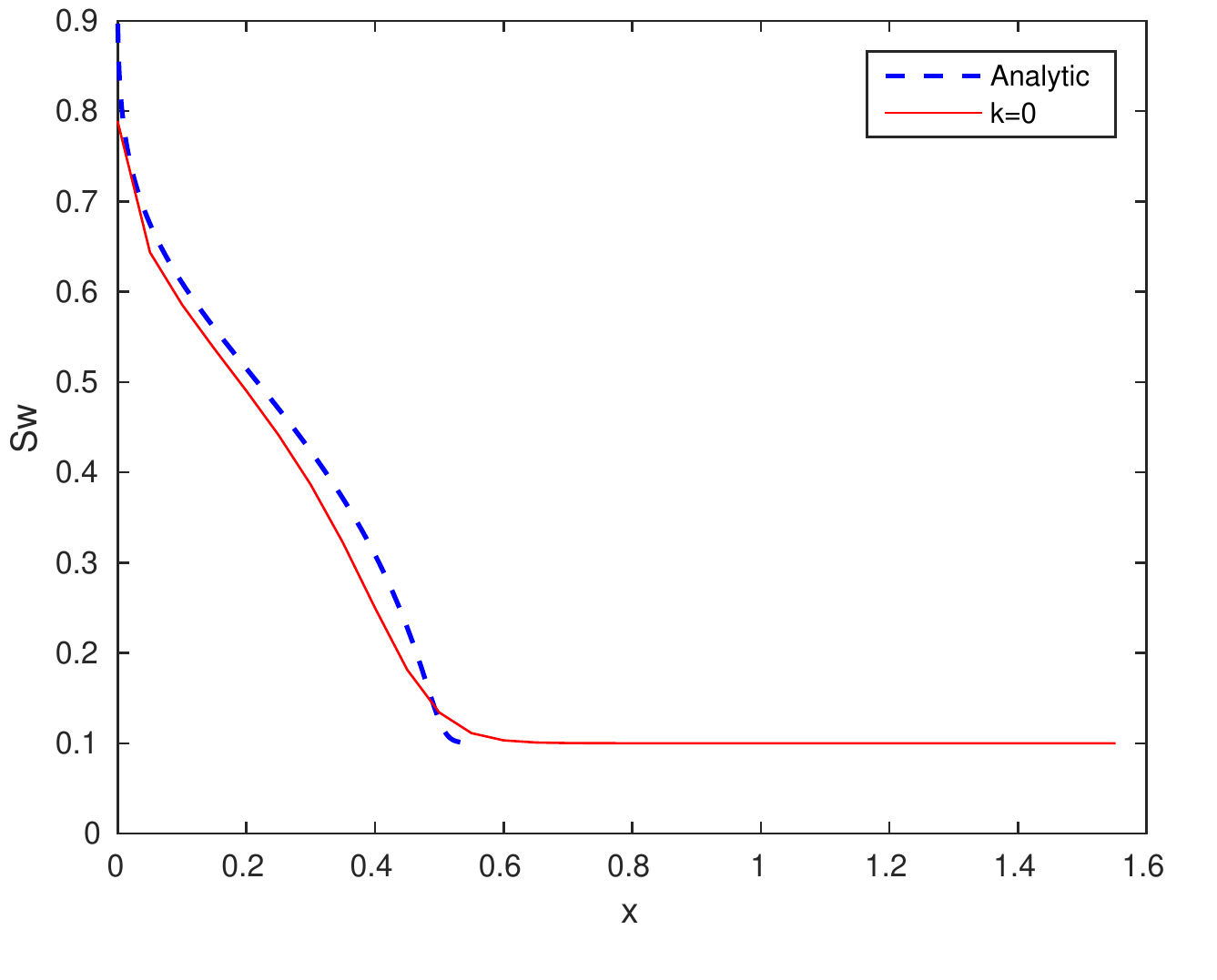}
    }
    \subfigure[\small{$k=1$}]{
        \includegraphics[trim = 40mm 80mm 40mm 90mm, clip, scale = 0.4]{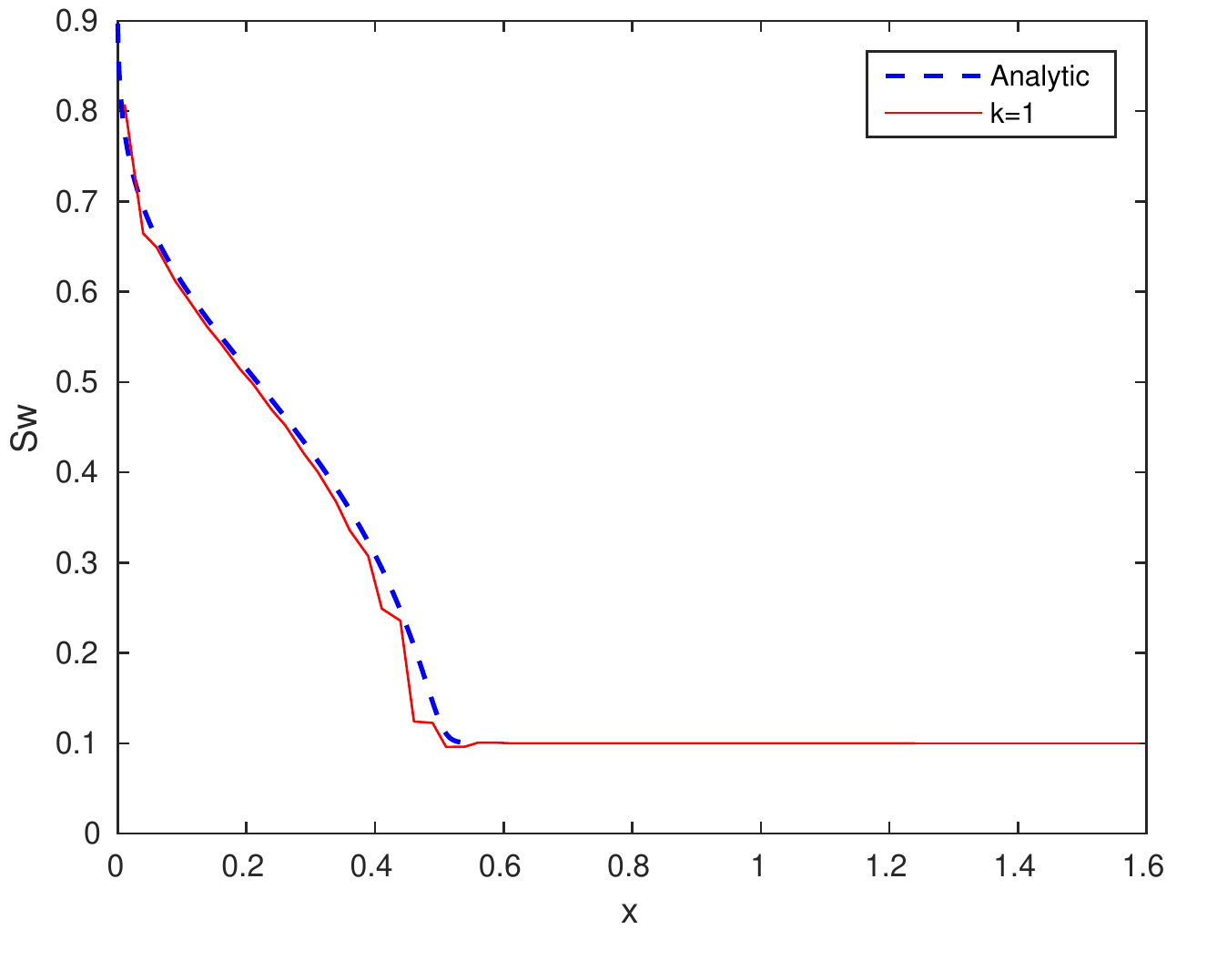}
    }    
    \subfigure[\small{$k=4$}]{
        \includegraphics[trim = 40mm 80mm 40mm 90mm, clip, scale = 0.4]{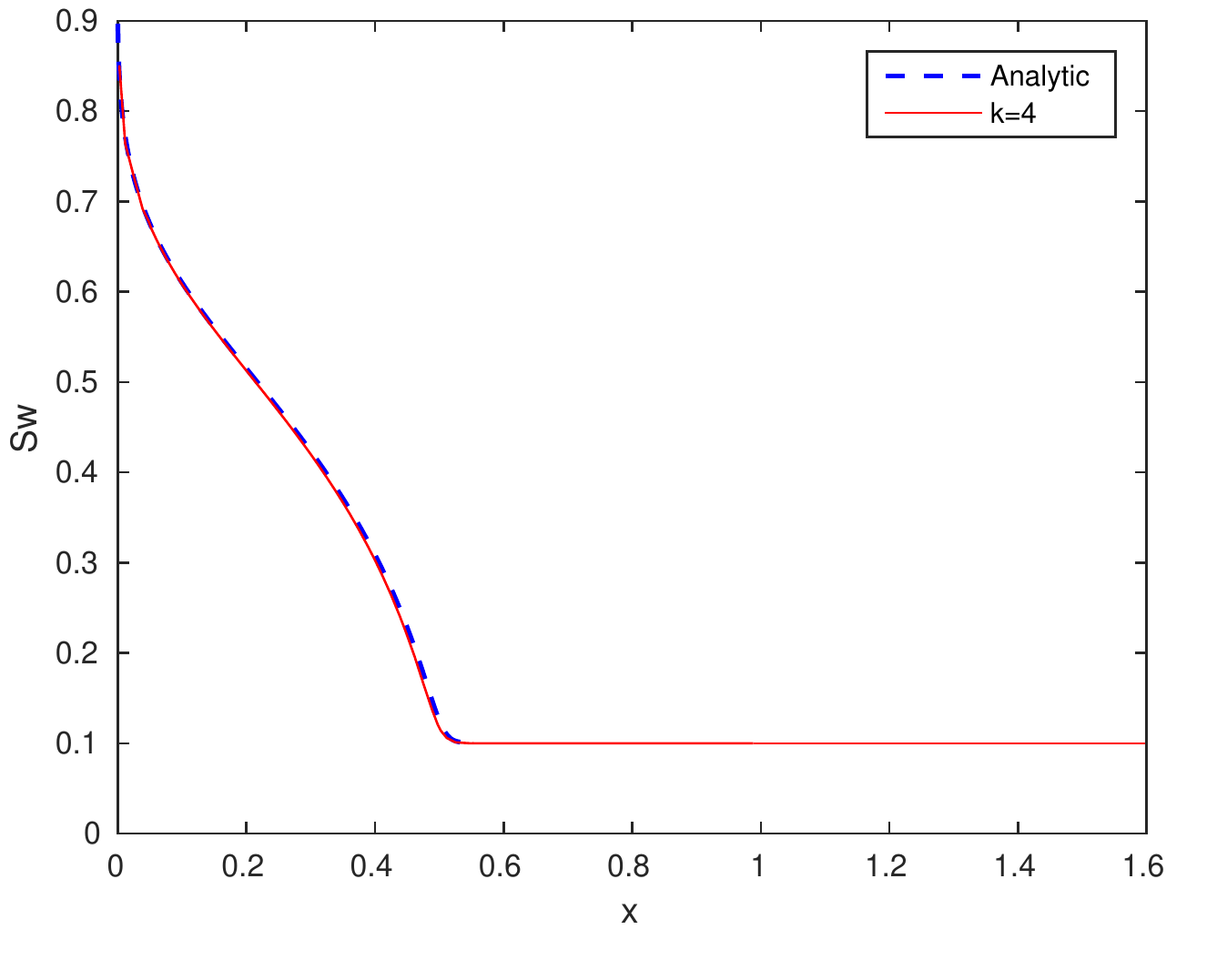}
    }     
\captionsetup{justification=justified}
\caption{Polynomial order refinement study.  Clearly as we increase the polynomial order the solution converges.}
\label{fig:mcWhorter_p_0}
\end{figure}

\begin{figure}[ht!]
\centering
\includegraphics[trim = 10mm 85mm 20mm 85mm, clip, scale = 0.5]{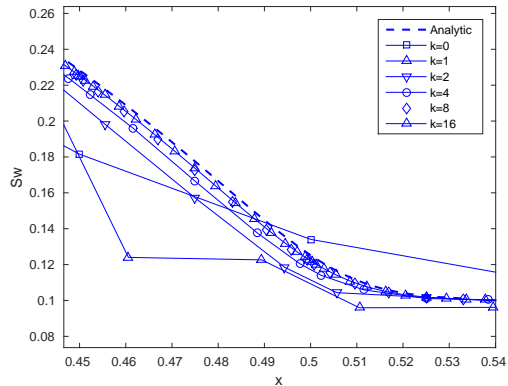} 
\caption{Zoom-in view of saturation profiles for polynomial orders from $k=0$ to $k=16$ for the McWhorter problem.}
\label{fig:mcworther18}
\end{figure}

\begin{figure}[ht!]
\centering
    \captionsetup{justification=centering}
    \subfigure[\small{$k=0$}]{
        \includegraphics[trim = 40mm 80mm 40mm 90mm, clip, scale = 0.5]{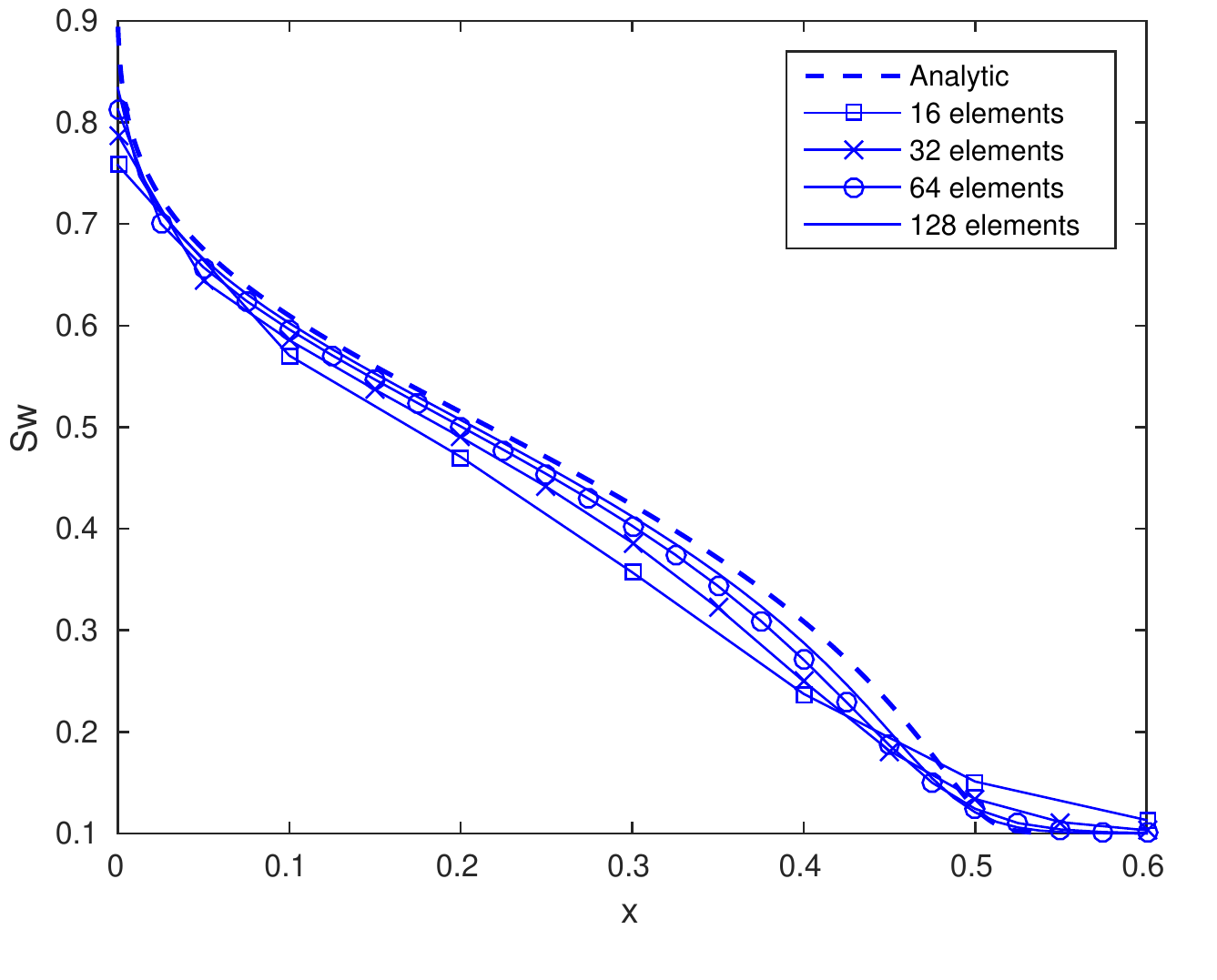}
        \label{fig:mcWhorter_h_k0}
    }
    \subfigure[\small{$k=1$}]{
        \includegraphics[trim = 40mm 80mm 40mm 90mm, clip, scale = 0.5]{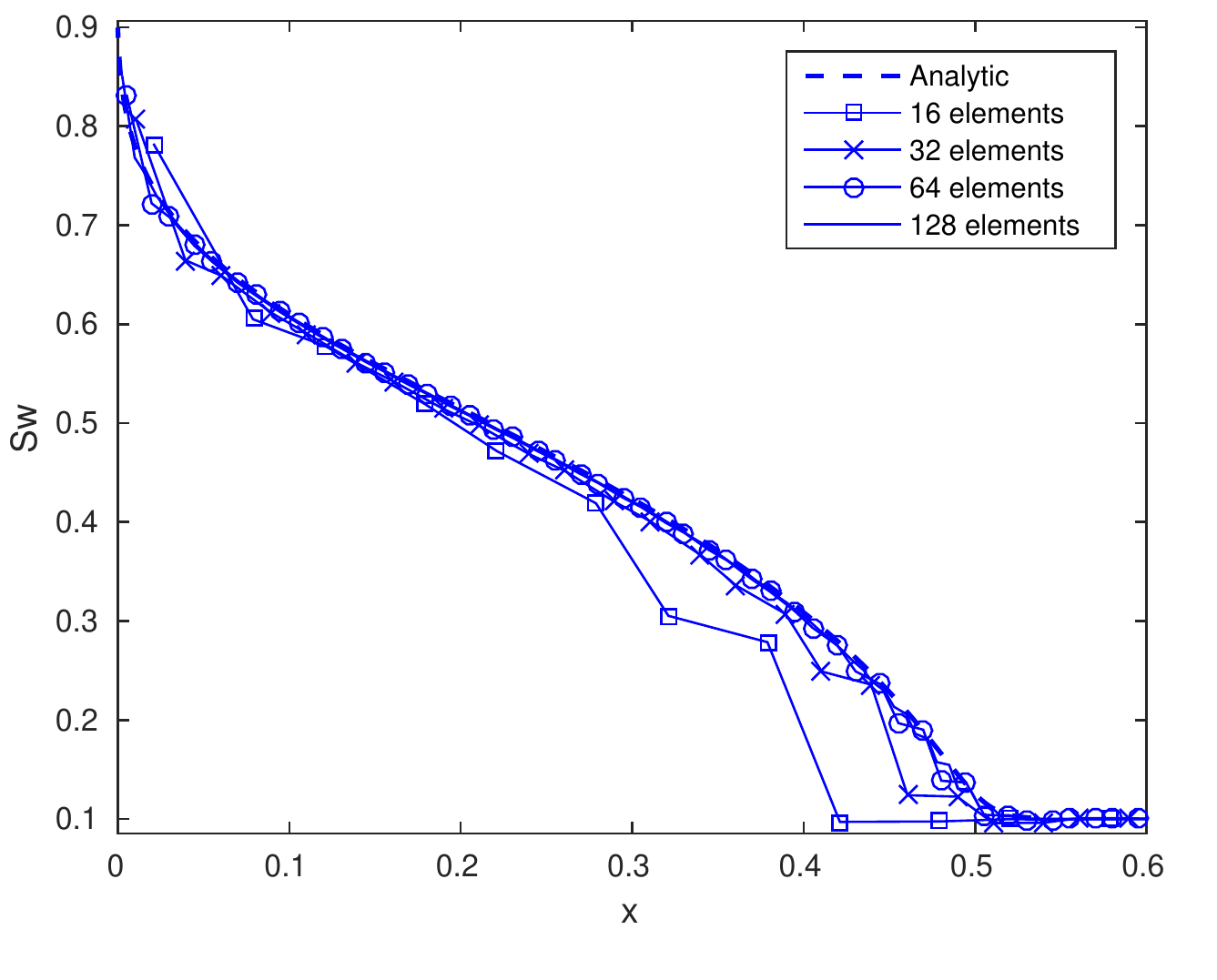}
    }    
    \\
    \subfigure[\small{$k=4$}]{
        \includegraphics[trim = 40mm 80mm 40mm 90mm, clip, scale = 0.5]{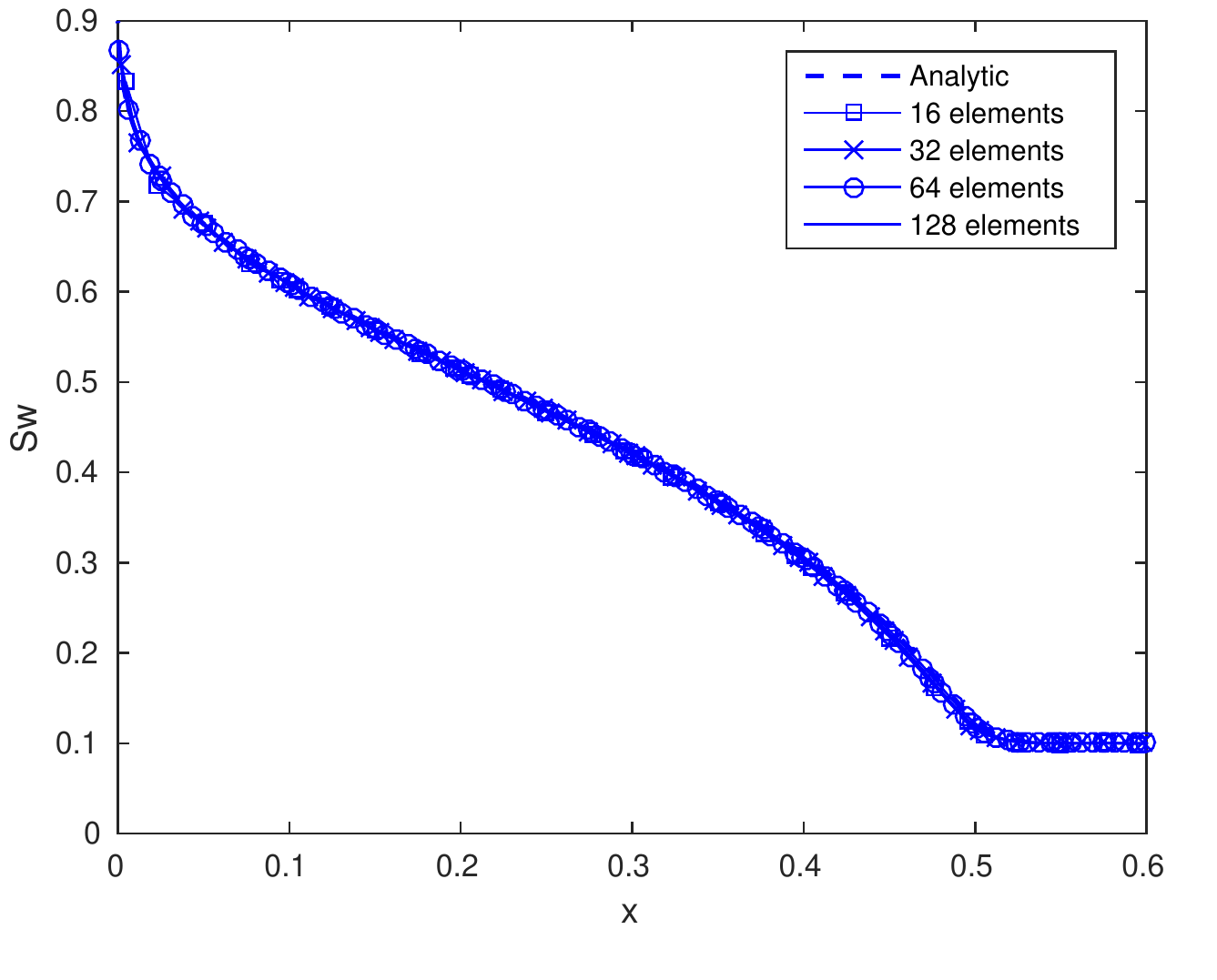}
    }     
    \subfigure[\small{$k=8$}]{
        \includegraphics[trim = 40mm 80mm 40mm 90mm, clip, scale = 0.5]{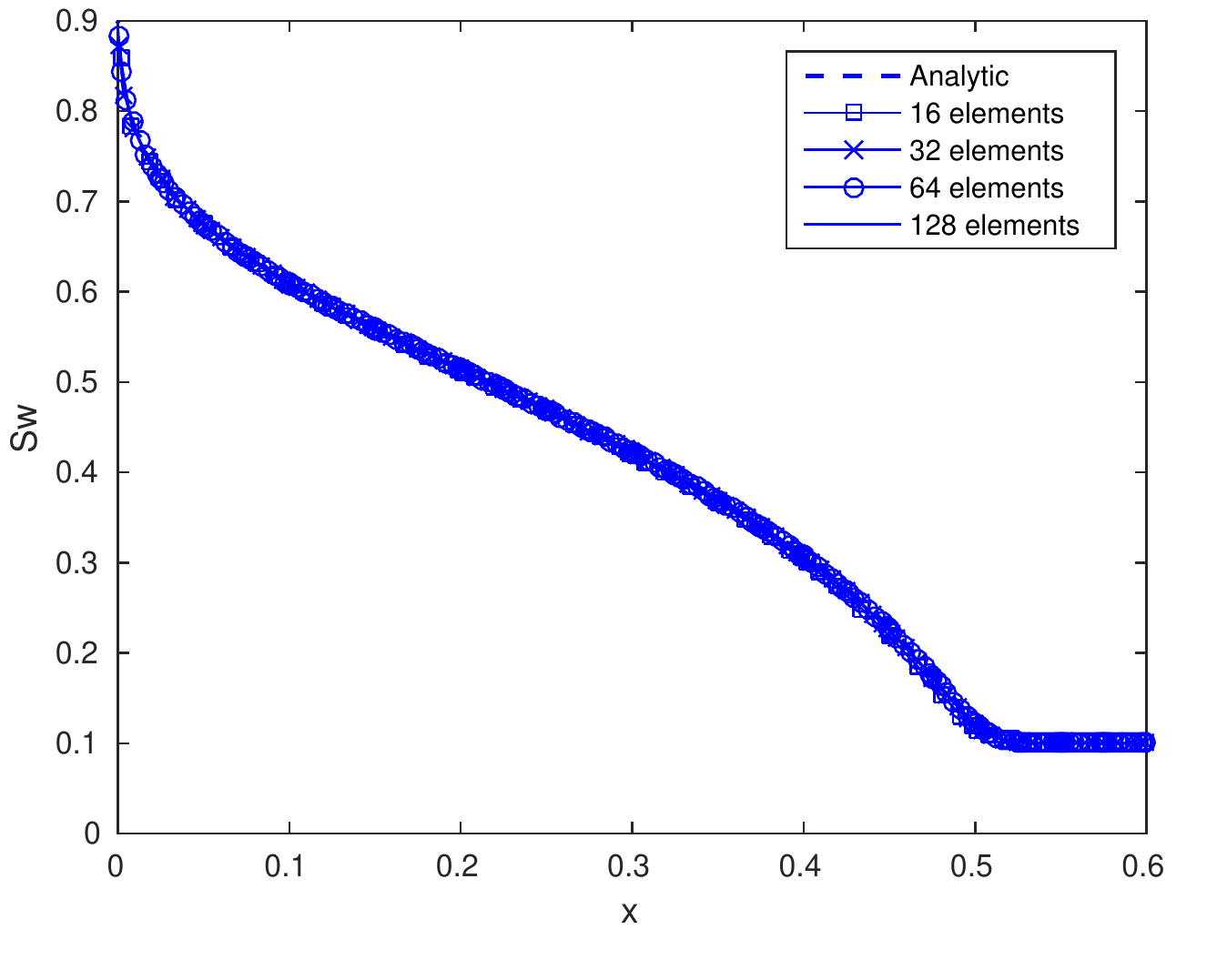}
    }      
\captionsetup{justification=justified}
\caption{Mesh refinement study for different polynomial degrees.  As we refine the mesh or increase the polynomial order the solution converges.}
\label{fig:mcworther-href}
\end{figure}

 The McWhorter problem is a one dimensional example of counter-current two-phase flow where capillary forces are present.  The governing equation is nonlinear, parabolic, and may be degenerate in the total velocity:
 \begin{equation}
 \Phi \frac{\partial s_w}{\partial t} 
 -
 \frac{\partial }{ \partial x}  \bigg( \frac{\lambda_w(s_w) \lambda_o(s_w)}{\lambda_w(s_w)+\lambda_o(s_w)}  |p_c'|  K\frac{\partial s_w}{\partial x} \bigg)
 =
 0, \quad \mbox{in}\quad \Omega \times (0,T),
 \end{equation}
 where we invoke Brooks--Corey relative permeabilities and capillary pressure:
  \begin{equation}
  \lambda_w(s_w) = \frac{s_w^{4}}{\mu_w}, \quad
  \lambda_o(s_w) = \frac{(1-s_w)^2 ( 1 - s_w^2)}{\mu_o},
\quad
  p_c(s_w) = p_d s_w^{-1/2}.
  \end{equation}
  A semi-analytical solution can be obtained for this problem (see \cite{mcwhorter1990exact}).  We fix the following parameters: entry pressure  $p_d= 5000$ (Pa), porosity $\Phi=0.3$, permeability $K= 10^{-8}$ (m$^2$), viscosities $\mu_o=\mu_w=10^{-3}$ (Pa.s). The domain is the interval $(0,1.6)$. At the left and right boundaries we prescribe $s_{\mathrm{D}} = 0.9$ and $s_{\mathrm{D}} = 0.1$ respectively, and the initial condition is taken as $s_w^0  = 0.1$.  The HDG method is used to discretize the problem in space, and implicit Euler is utilized in time.  We use implicit time stepping as the McWhorter problem is parabolic, and explicit time marching schemes have a severe time step restriction for this class of problems.  In one dimension, static condensation for the HDG method always results in a matrix that is tridiagonal, for \textit{any} polynomial order $k\ge0$, since the intersection of two adjacent elements is a single point.
\\[12pt]
In Fig.~\ref{fig:mcWhorter_profile} wetting phase saturation profiles at different times are displayed using a high order approximation of $k=8$ on a mesh with 32 elements.  Overshoot and undershoot remain bounded due to the use of Newton's method.  A polynomial refinement study is performed in Fig.~\ref{fig:mcWhorter_p_0}.  A relatively coarse mesh of 64 elements is used and we run the simulation to $T=80$.  The polynomial order varies:  $k=0, 1$ and $k=4$.  We note that the saturation front for the case $k=0$ lags behind the front of the semi-analytical solution. As the polynomial order increases, so does the accuracy.  We show in Fig.~\ref{fig:mcworther18} a zoom-in view of the saturation profile in the neighborhood of $x=0.5$ for a geometric sequence of polynomial orders, $k\in\{0,1,2,4,8,16\}$.  We observe convergence of the solution with polynomial degree refinement.
Even though the solution has poor regularity, high order polynomial approximations bestow an advantage without any use of slope limiting.
\noindent  Similarly, if we fix the polynomial order and solve the problem on successively refined meshes, we obtain improved accuracy in the numerical solution. Fig.~\ref{fig:mcworther-href} shows the saturation profiles on meshes with 16, 32, 64 and 128 elements. The polynomial approximation varies: $k\in \{0,1, 2, 4\}$.
We observe that the piecewise constant ($k=0$) solution gives a poor approximation in comparison to the high order polynomials; there is a visible gap around $x=0.4$ between the piecewise constant approximation and the analytic solution.  The figure shows that as we refine the mesh, the approximation does improve.  However, to match the accuracy of the higher order approximations, one would have to use a mesh that is very fine.  On a mesh with 128 elements, the picewise constant solution is visibly of lesser quality when compared to the piecewise quartic approximation on a coarse mesh of 16 elements.
\\[12pt]

\begin{figure}[ht!] 
\centering
        \includegraphics[trim = 40mm 80mm 40mm 90mm, clip, scale = 0.5]{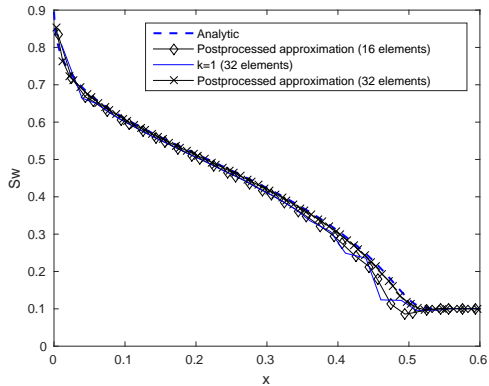}
\caption{Effect of postprocessing $k=1$ approximation on a mesh with 16 elements and 32 elements, compared to the $k=1$ approximation on a mesh with 32 elements.  The spurious oscillations are eliminated, and a much coarser mesh may be used instead of mesh refinement.}
\label{mcWhorter_p_00}
\end{figure}

\noindent  The HDG method brings with it the ability to apply a simple postprocessing to the wetting phase saturation $s_{wh}$ to obtain a new approximation $s_{wh}^*$ that converges at a rate of $k + 2$ in the $L^2$-norm.  As discussed in Section~\ref{sec:postproc}, the postprocessing takes the form of a simple diffusion problem that is element local, as such it is completely data parallel.  In~\cite{fabien2} a work-precision study is performed which illustrates that the postprocessing is cheaper than solving a refined globally coupled problem.  Fig.~\ref{mcWhorter_p_00} shows that the local postprocessing removes the spurious oscillations that occur near $x=0.5$.  We note that even though the oscillations remain bounded, mesh refinement is not as effective as the postprocessing with respect to eliminating the oscillations, see Fig.~\ref{fig:mcworther-href}.  In other words, the cheap postprocessing may be useful in situations where a more accurate solution is desired without further mesh refinement.  We point out that the postprocessing is able to recover an enhanced approximation using an even coarser mesh of 16 elements (Fig.~\ref{mcWhorter_p_00}).

\subsection{Buckely--Leverett problem}
 \label{sec:ex_BL}
\begin{figure}[ht!]
\centering
\includegraphics[trim = 10mm 80mm 20mm 85mm, clip, scale = 0.45]{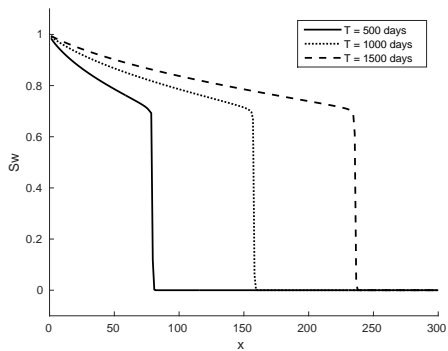} 
\caption{Saturation profiles at different times for the Buckley-Leverett problem using piecewise linears and a mesh of 256 elements.}
\label{fig:buck_k1}
\end{figure}

\begin{figure}[ht!]
\centering
    \captionsetup{justification=centering}
    \subfigure[\small{$k=0$}]{
        \includegraphics[trim = 40mm 80mm 40mm 90mm, clip, scale = 0.5]{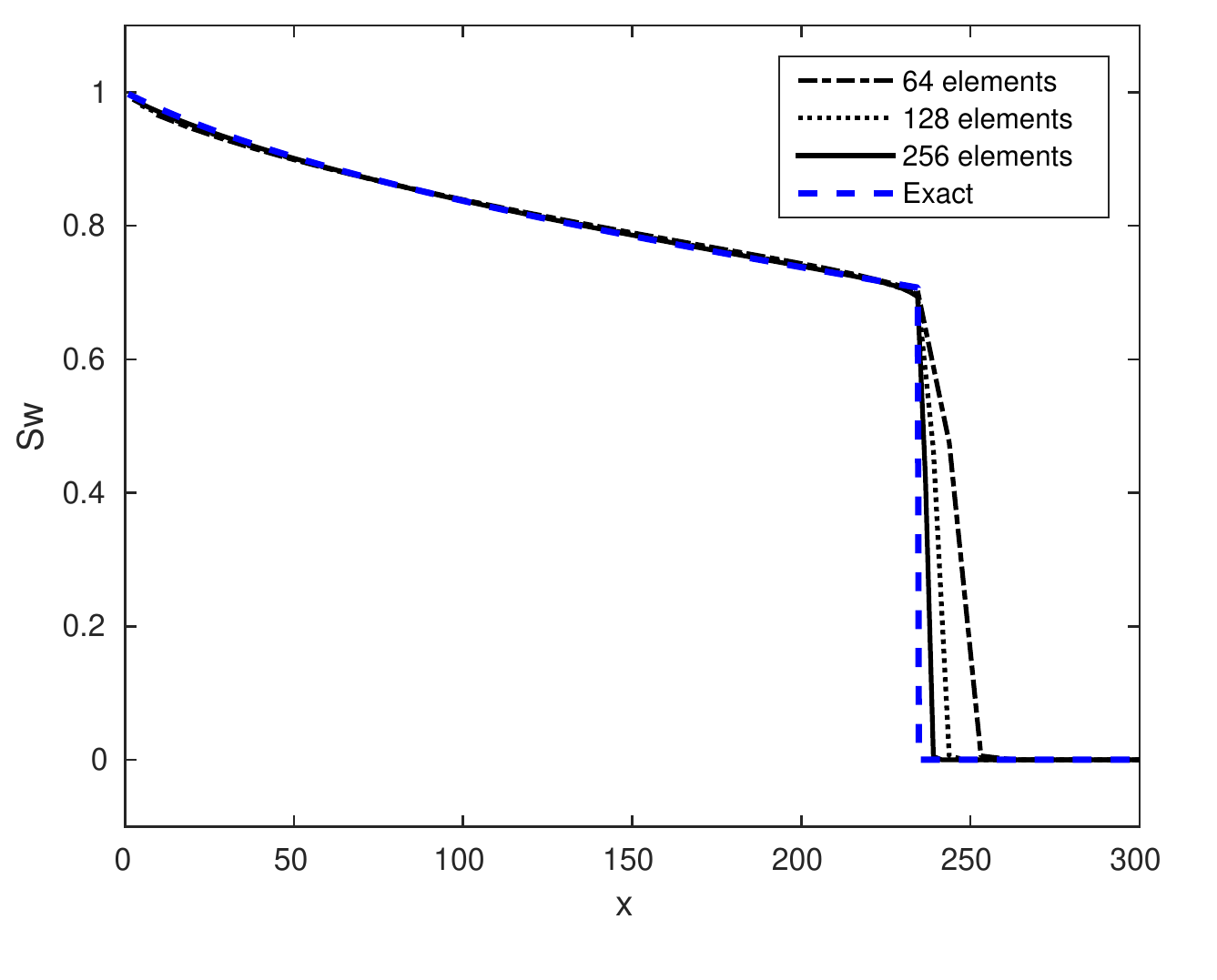}
    }
    \subfigure[\small{$k=1$}]{
        \includegraphics[trim = 40mm 80mm 40mm 90mm, clip, scale = 0.5]{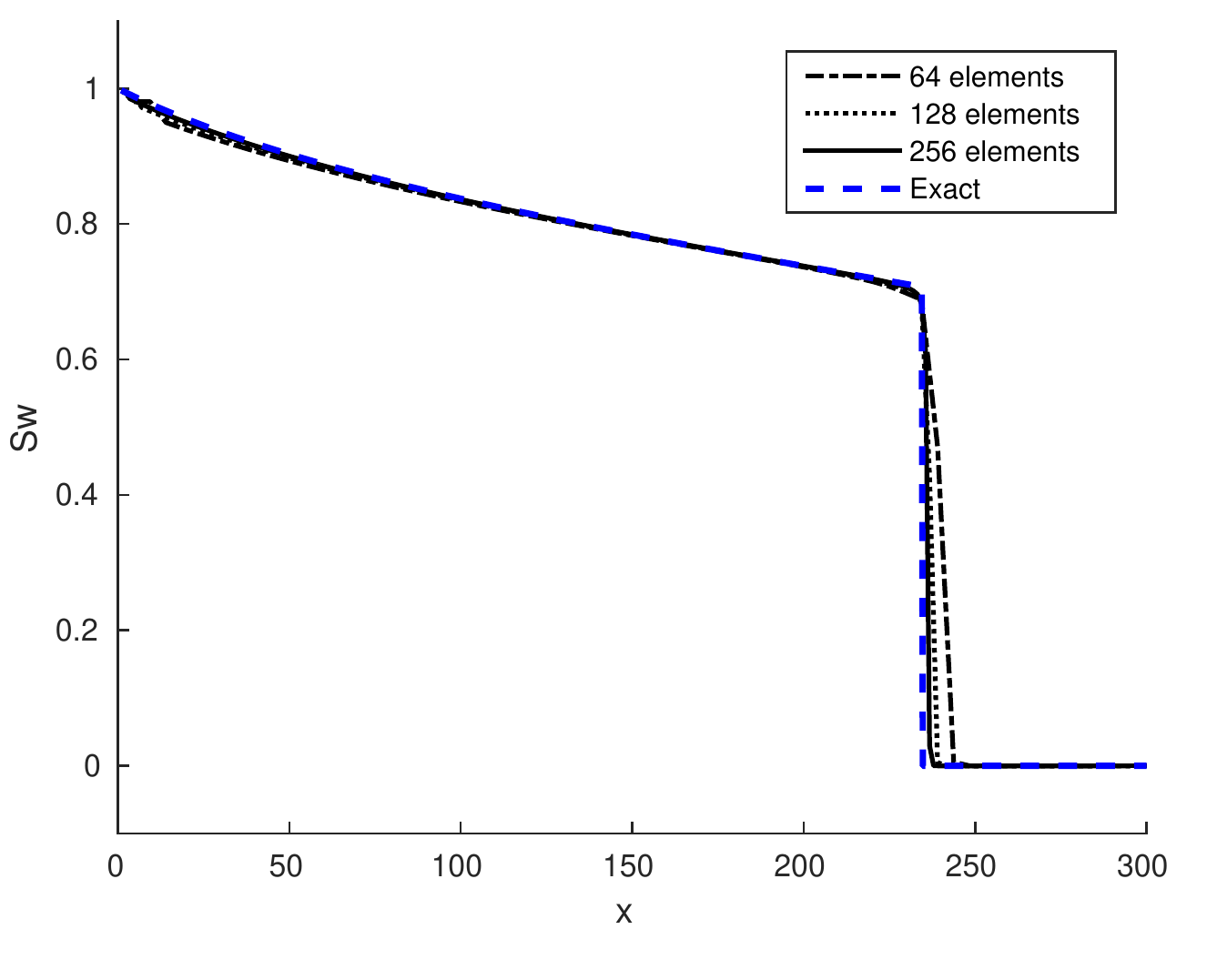}
    }    
    \\
    \subfigure[\small{$k=1$ (zoom-in)}]{
        \includegraphics[trim = 40mm 80mm 40mm 90mm, clip, scale = 0.5]{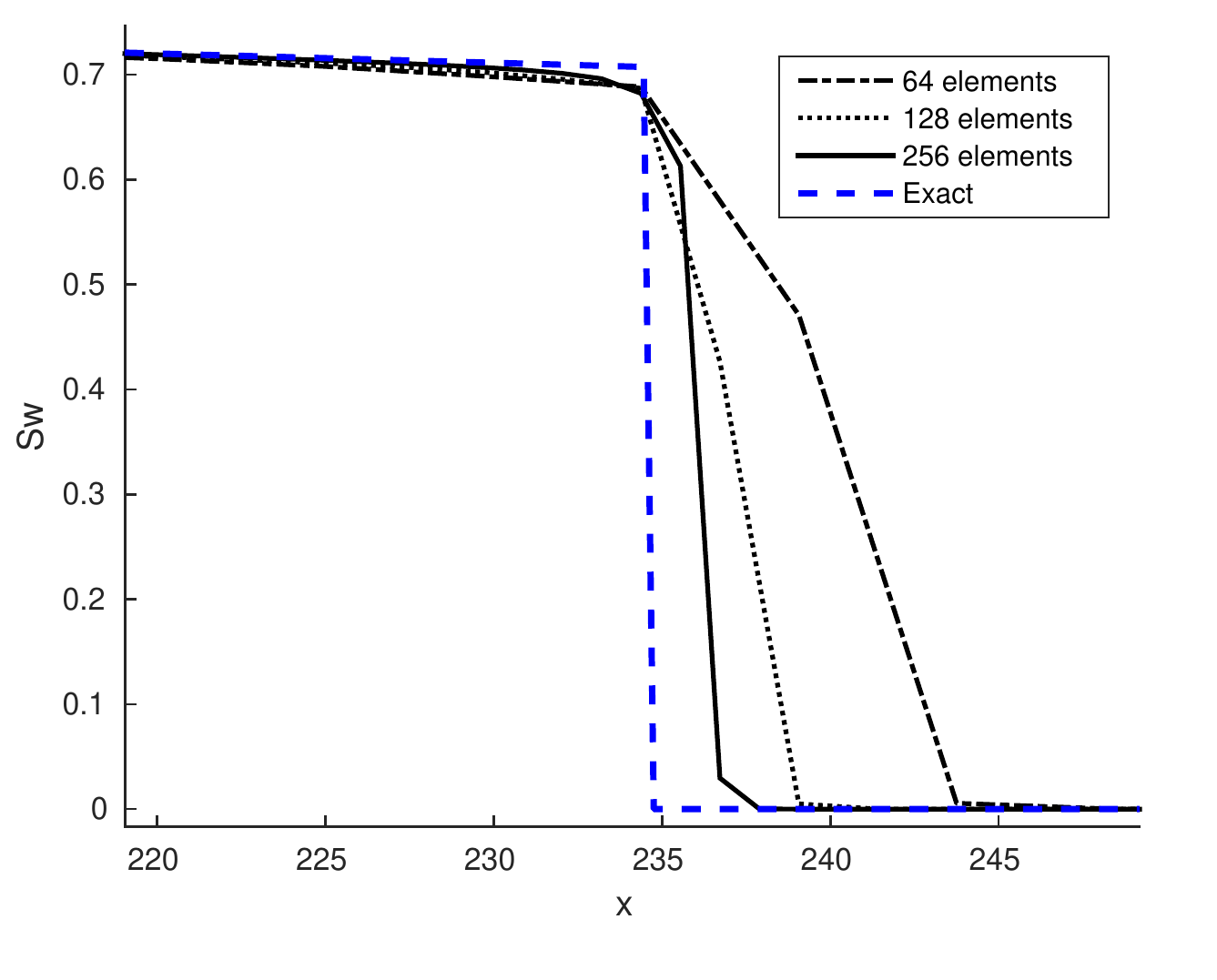}
        \label{fig:buck_zoom}
    }     
\captionsetup{justification=justified}
\caption{Buckely–-Leverett problem.}
\label{fig:buck_href}
\end{figure}

The Buckely--Leverett problem is a popular one-dimensional model problem that is used to validate numerical methods for two-phase flows with zero capillary pressure \cite{buckley1942mechanism}.  The equations~\eqref{eq1}-\eqref{eq3} simplify to:
\begin{equation}
\Phi
\frac{\partial s_w}{\partial t}
+
\frac{\partial }{\partial x}
\left(\frac{\lambda_w(s_w)}{\lambda_w(s_w)+\lambda_o(s_w)} u_t\right)
= 0.
\label{buck_eq}
\end{equation}
The computational domain is taken to be $\Omega = (0,300)$ (m).  The relative permeabilities are the same as in Section~\ref{sec:ex_Mc}.  The porosity is set to $\Phi = 0.2$, the velocity $u=3\cdot 10^{-7}$ (m/s) and viscosities $\mu_w=\mu_n=1$ (Pa s).  The initial saturation is equal to zero and the Dirichlet boundary condition at the left boundary, $x=0$, is $s_{\mathrm{D}} = 1$.
As equation~(\ref{buck_eq}) is hyperbolic, we use the standard fourth order Runge-Kutta explicit time stepping.  This choice is made as implicit methods may add a large amount of diffusion smearing the concentration front.  The selected initial condition propagates into a solution that develops a shock, and higher order discontinuous Galerkin methods require slope limiting to handle overshoot and undershoot \cite{cockburn1998runge}.  We employ a standard minmod limiter~\cite{roe1986characteristic}.  The time step is chosen to satisfy a CFL condition, $\Delta t  = 0.5 \Phi h/u_t$, which yields $\Delta t = 1.929$ days. The polynomial degree is set equal to one.  The saturation profile is displayed in Fig.~\ref{fig:buck_k1} for $t=500,$ $1000,$ and $1500$ days.  
We then perform a mesh refinement study and fix the final time $T=1500$ days. Fig.~\ref{fig:buck_href} shows the saturation profiles using either piecewise constant or piecewise linear approximations on successively refined meshes. We also include a zoom-in view for the piecewise linear solution in Fig.~\ref{fig:buck_zoom}. As expected, accuracy is increased as we refine the mesh and increase the polynomial degree.

\subsection{Convergence rates}
\label{sec:ex_Manu}
\begin{figure}[ht!]
\centering
\begin{minipage}{0.45\textwidth}
\centering
\hspace*{-15ex}
\includegraphics[trim = 10mm 80mm 20mm 85mm, clip, scale = 0.55]{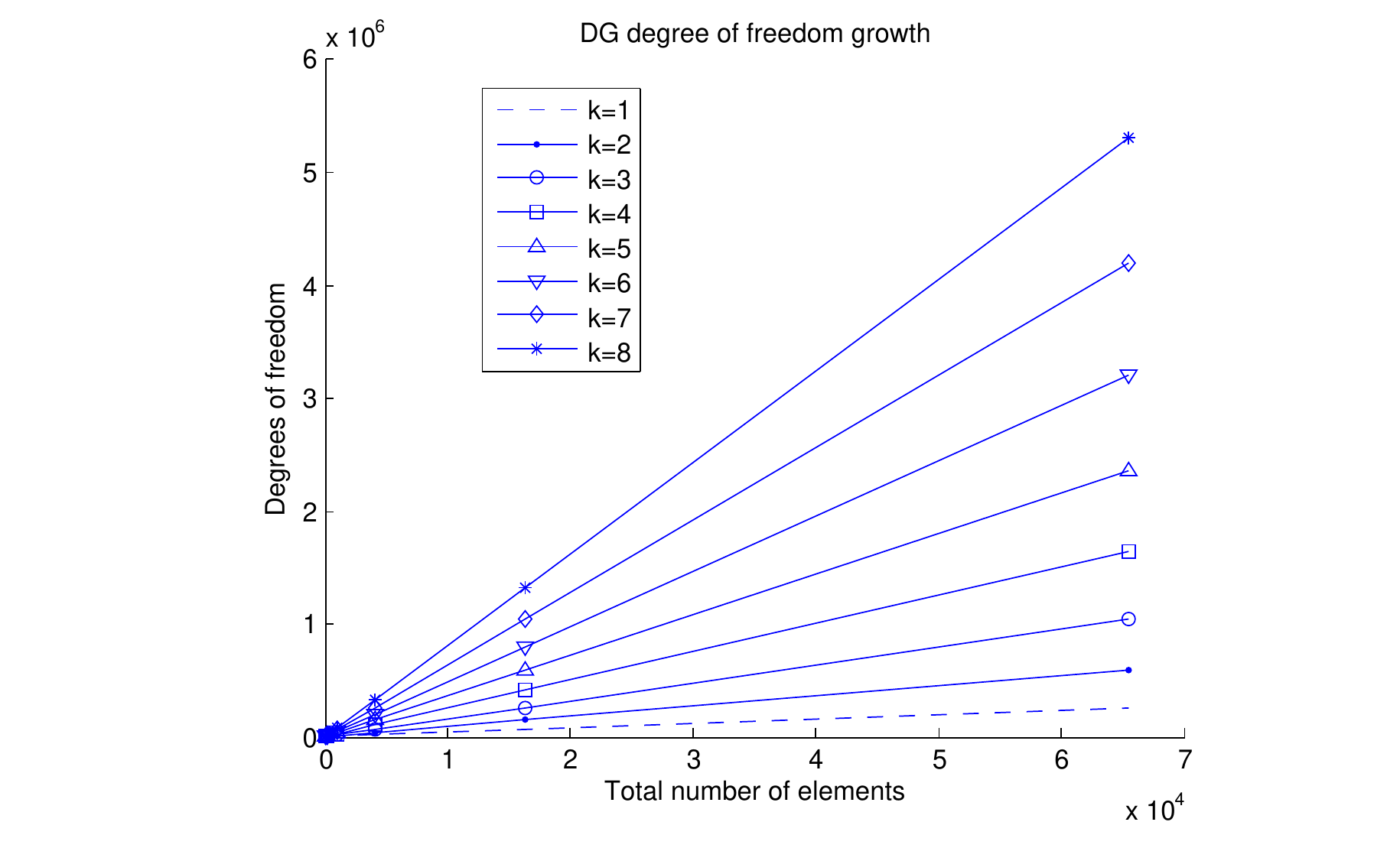} 
\caption{DG degrees of freedom}
\label{dof_1}
\end{minipage}
\begin{minipage}{0.45\textwidth}
\centering
\hspace*{-5ex}
\includegraphics[trim = 10mm 80mm 20mm 85mm, clip, scale = 0.55]{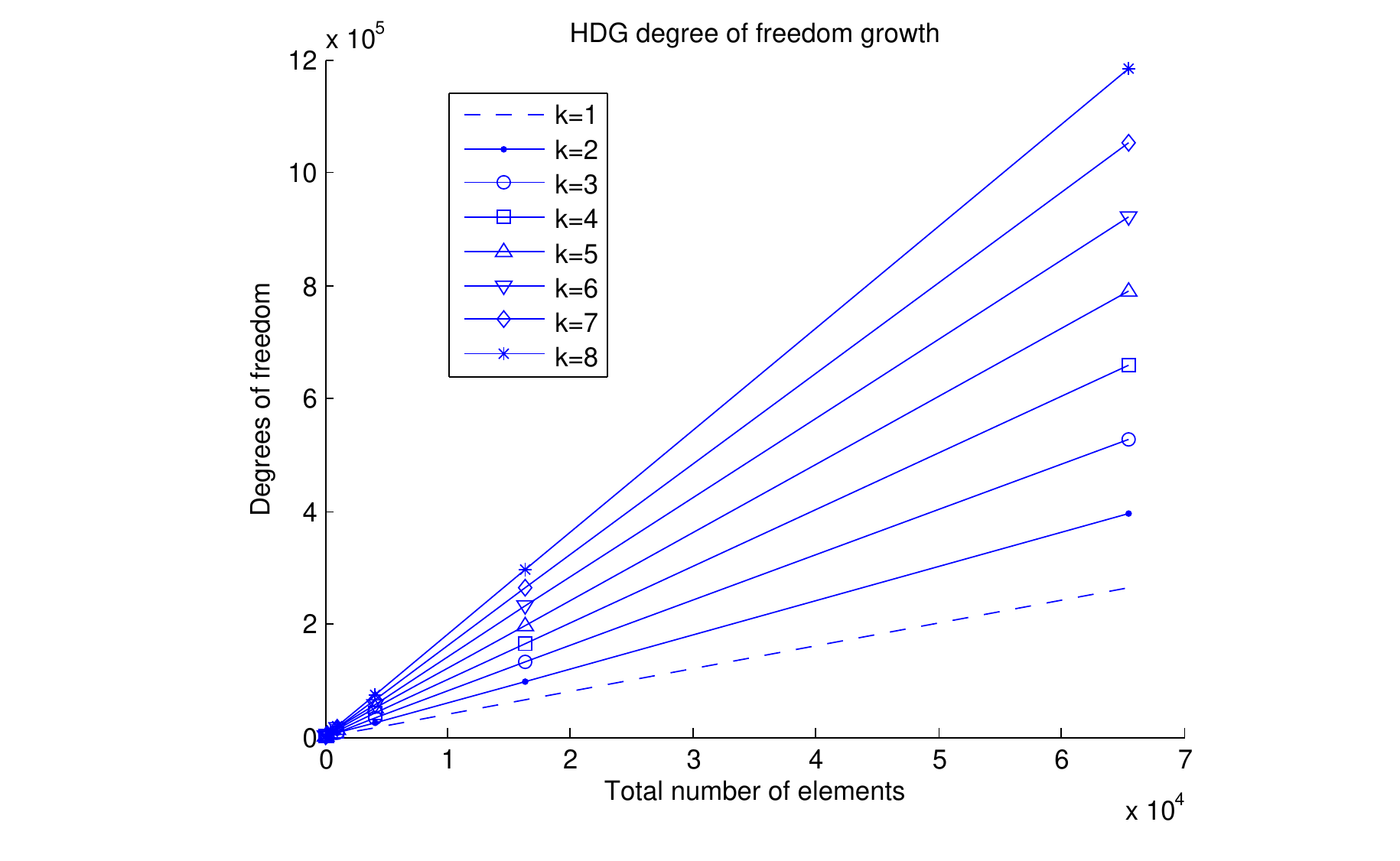} 
\caption{HDG degrees of freedom}
\label{dof_2}
\end{minipage} 
\caption{(Left)  Classical DG degree of freedom growth.  (Right) HDG degree of freedom growth.}
\label{dof_00}
\end{figure}

\begin{figure}[htb!]
\centering
\includegraphics[trim = 10mm 80mm 20mm 85mm, clip, scale = 0.6]{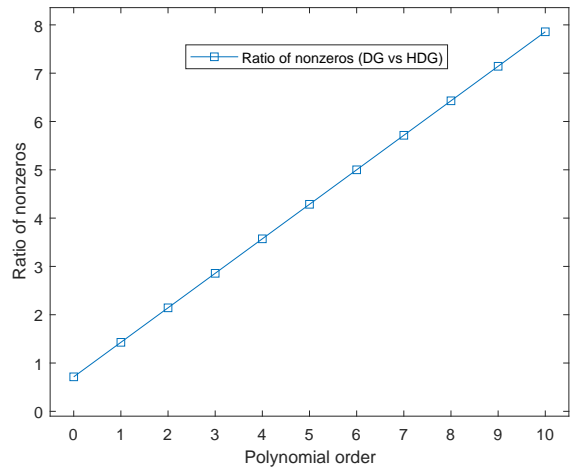} 
\caption{Ratio of nonzeros in discretization matrix for DG vs HDG.}
\label{dof_11}
\end{figure}

We take $\Omega$ to be the unit square, and consider the following manufactured solutions:
\begin{equation}
\begin{split}
            s_w &= 0.5(t+1) - (7.0) x y (1-x) (1-y) e^{(-x^2-y^2)}, 
            \\
            p_w &= (t+1) + x y \tanh(1-x) \tanh(1-y).
\end{split}
\end{equation}
The function $s_w$ is designed so that $0<s_w<1$.  Dirichlet boundary conditions are imposed for both wetting phase pressure and wetting phase saturation.  Capillary pressure and relative permeabilities are defined in Section~\ref{sec:ex_Mc}, with entry pressure $p_d = 1000$. The viscosities are $\mu_o = 10^{-2}$ and $\mu_w = 10^{-3}$.
For simplicity, the medium is assumed to be homogeneous, with permeability $ K = 10^{-4}$. The mesh is made of $N\times N$ square elements.  The time-step is chosen as $\Delta t = 1/(N^{k+1})$, and the final time is $T=1$.  In Table~\ref{tab1_sw} we verify that the numerical scheme  generates approximations to the saturation that agree with the expected convergence rates.  We vary the mesh sizes and the polynomial degree. We compute the errors in the $L^2$ norm for the saturation and its gradient, at the final time. We observe that $s_{wh}$ converges at the optimal rate of $\mathcal{O}(h^{k+1})$.  The HDG method also has the property that the gradient ${\bf q}_h$, also converges at the optimal rate of $\mathcal{O}(h^{k+1})$.  Primal DG methods require one to reconstruct the gradient from the variable $s_{wh}$,  which results in a loss of accuracy. The accuracy of ${\bf q}_h$ is important, because the gradient of capillary pressure in the Darcy flow requires the gradient of the wetting phase saturation.  We also show in Table~\ref{tab1_sw} the numerical rates obtained for the quantity
($s_{wh} - s_{wh}^*$) in the $L^2$ norm; convergence of order $\mathcal{O}(h^{k+2})$ is observed.  Convergence rates for the numerical scheme of the Darcy system~(\ref{eq1})-(\ref{eq2}) are presented in Table~\ref{tab1_pw}.  It is evident that \textit{both} the wetting phase pressure and total velocity converge at the optimal rate of $\mathcal{O}(h^{k+1})$ in the $L^2$ norm.  As our algorithm is semi-implicit, having an accurate total velocity is of critical importance as it is inserted into the saturation equation~(\ref{eq3}).
\\[12pt]
In Fig.~\ref{dof_00} we compare the degree of freedom growth for classical primal DG methods to that of HDG, in a log-log plot.  We assume a uniform Cartesian mesh in two dimension.  For piecewise quartic approximation, DG methods have roughly 2.5 times more unknowns than HDG.  Increasing the polynomial order to $k=8$, DG methods have roughly 4.5 times more unknowns than HDG.  
Fig.~\ref{dof_11} examines the total number of nonzero entries in the discretization matrix for DG compared to the statically condensed HDG.  For $k=5$, the DG method has about 4.3 times more nonzero entries as the HDG method.  This ratio increases to 6.4 for the case $k=8$.
\subsection{2D heterogeneous medium}
\label{section_het1}
\begin{figure}[ht!]
\centering
    \captionsetup{justification=centering}
   
    \subfigure[\small{$k=1$}]{
        \includegraphics[trim = 40mm 80mm 30mm 90mm, clip, scale = 0.5]{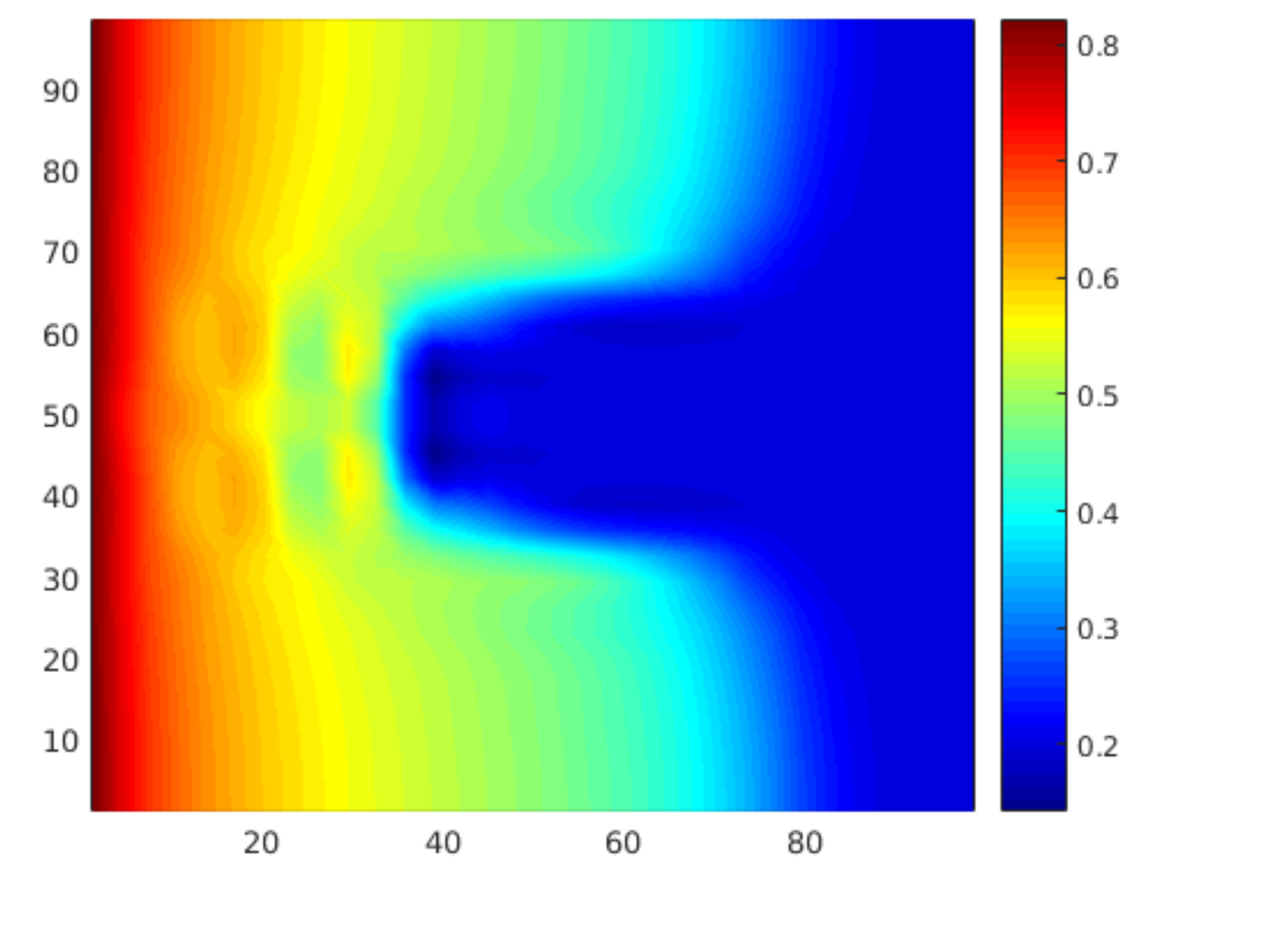}
    }
    \subfigure[\small{$k=2$}]{
        \includegraphics[trim = 40mm 80mm 30mm 90mm, clip, scale = 0.5]{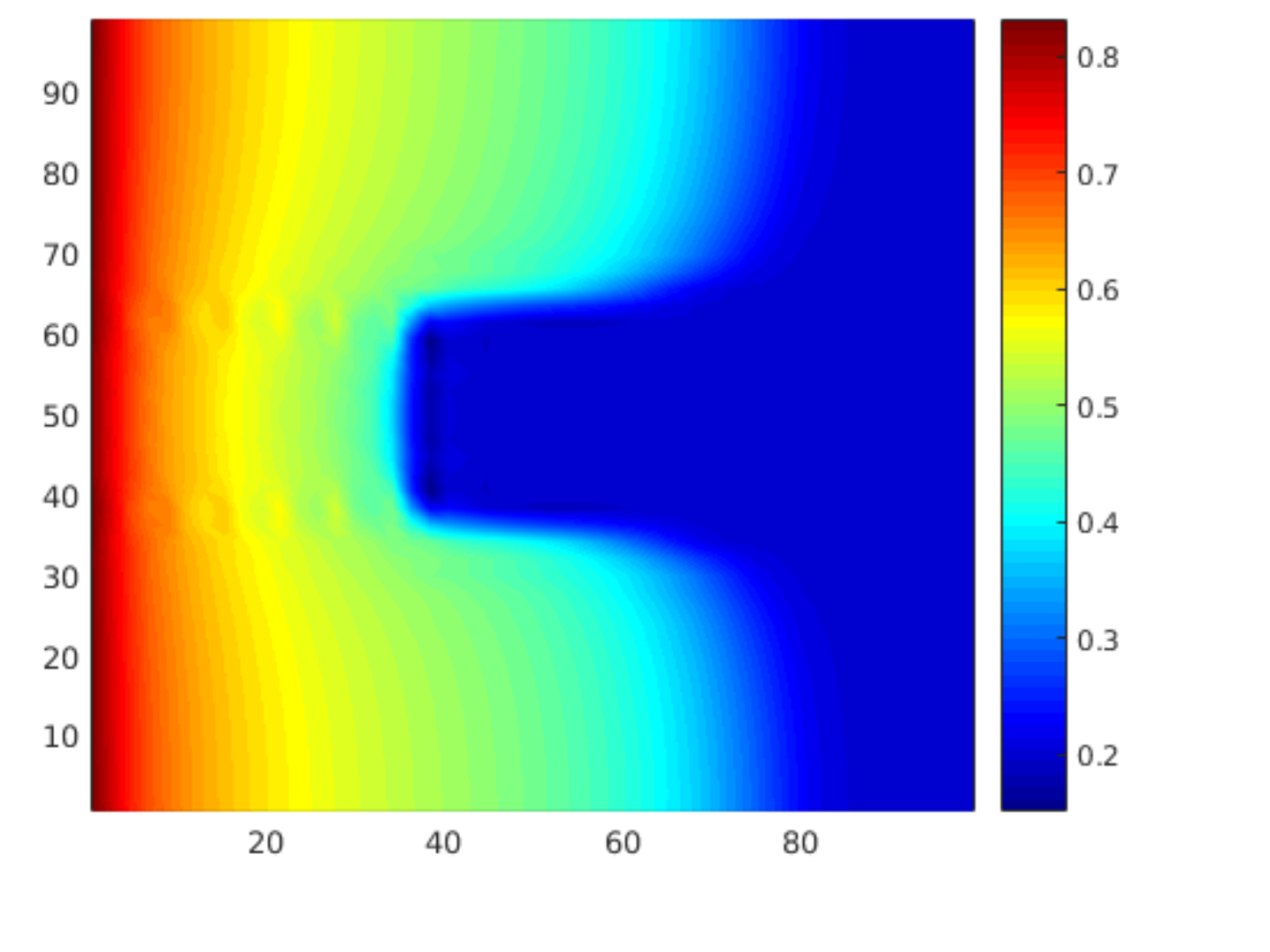}
    }    
    \\
    \subfigure[\small{$k=4$}]{
        \includegraphics[trim = 40mm 80mm 30mm 90mm, clip, scale = 0.5]{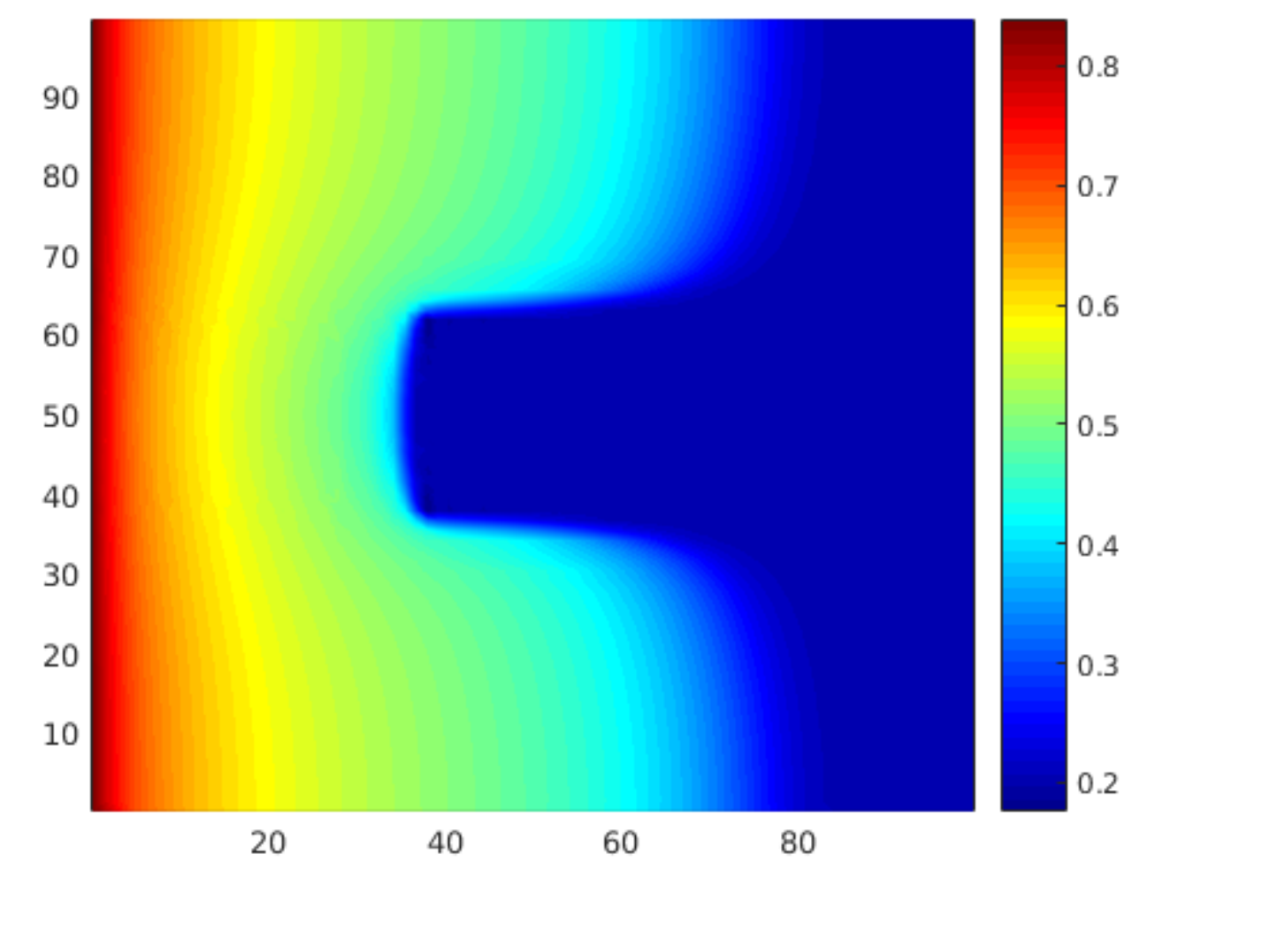}
    }
    \subfigure[\small{$k=8$}]{
        \includegraphics[trim = 40mm 80mm 30mm 90mm, clip, scale = 0.5]{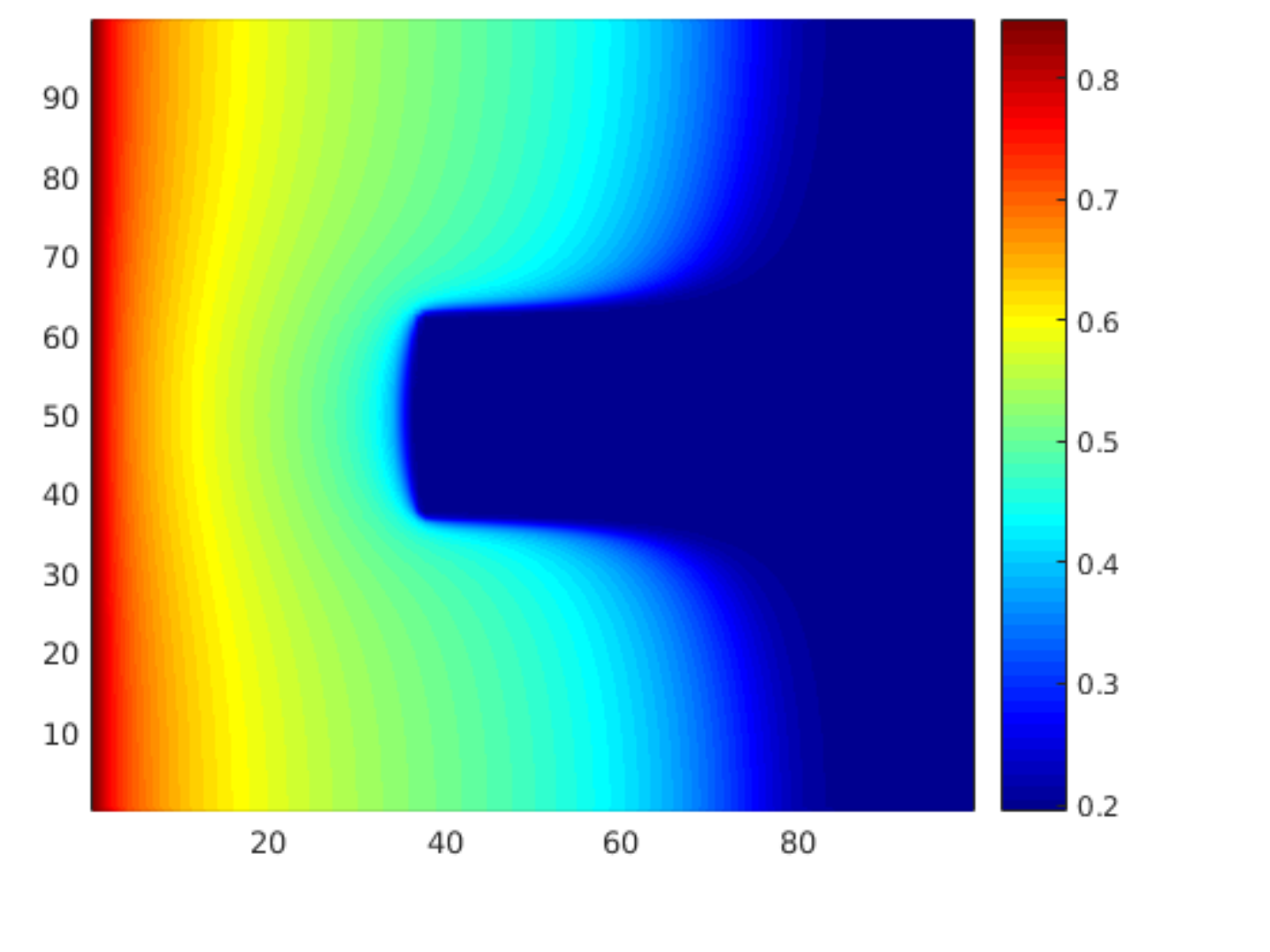}
    }      
\captionsetup{justification=justified}
\caption{Wetting phase saturation contour at $T=700$ days for different polynomial orders.  As the order increases, overshoot and undershoot diminish.}
\label{fig:utah_p_0}
\end{figure}

\begin{figure}[ht!]
\centering
\begin{minipage}{0.45\textwidth}
\centering
\hspace*{-25ex}
\includegraphics[trim = 10mm 80mm 20mm 85mm, clip, scale = 0.65]{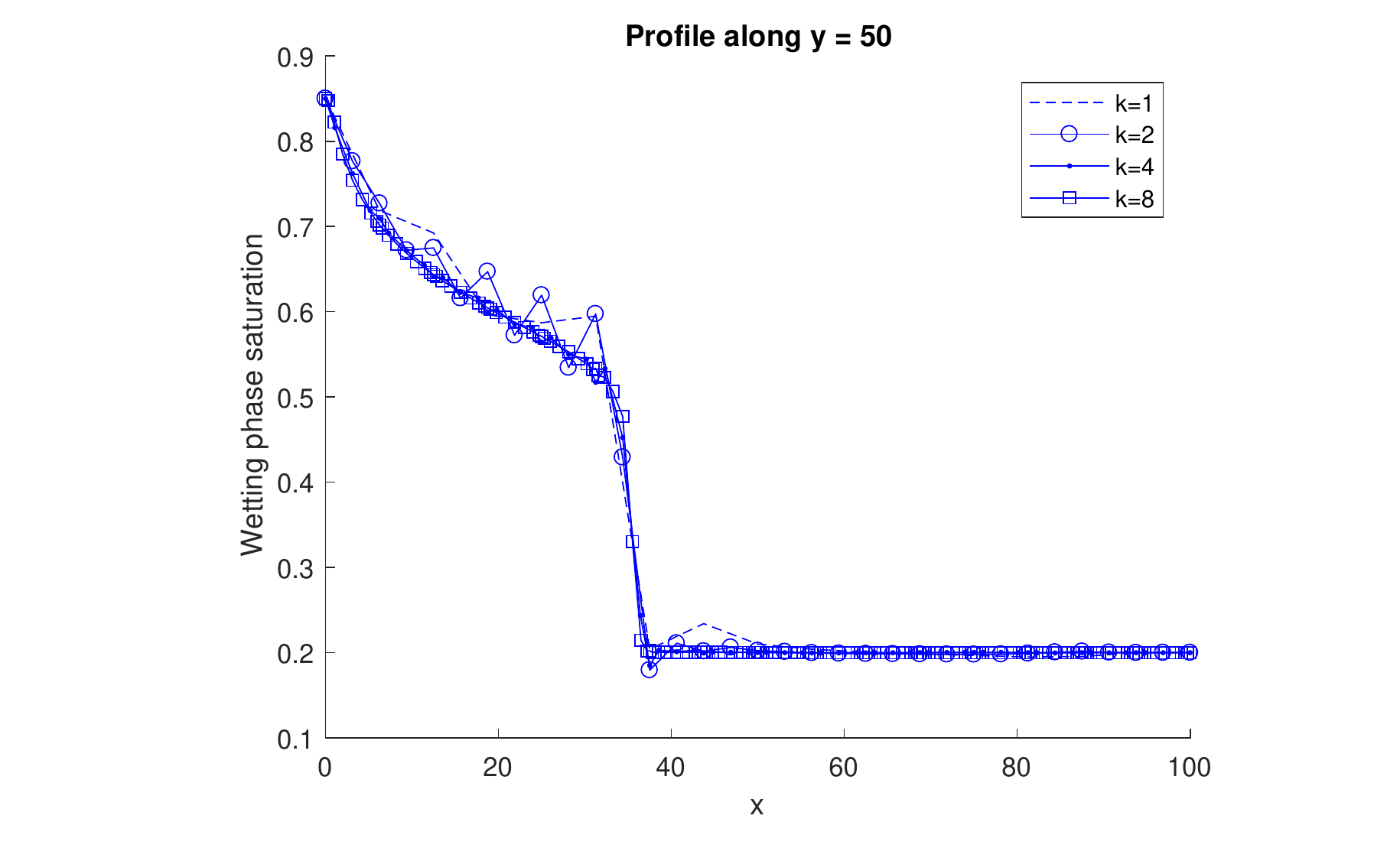}
\end{minipage}
\begin{minipage}{0.45\textwidth}
\centering
\hspace*{-15ex}
\includegraphics[trim = 10mm 80mm 20mm 85mm, clip, scale = 0.65]{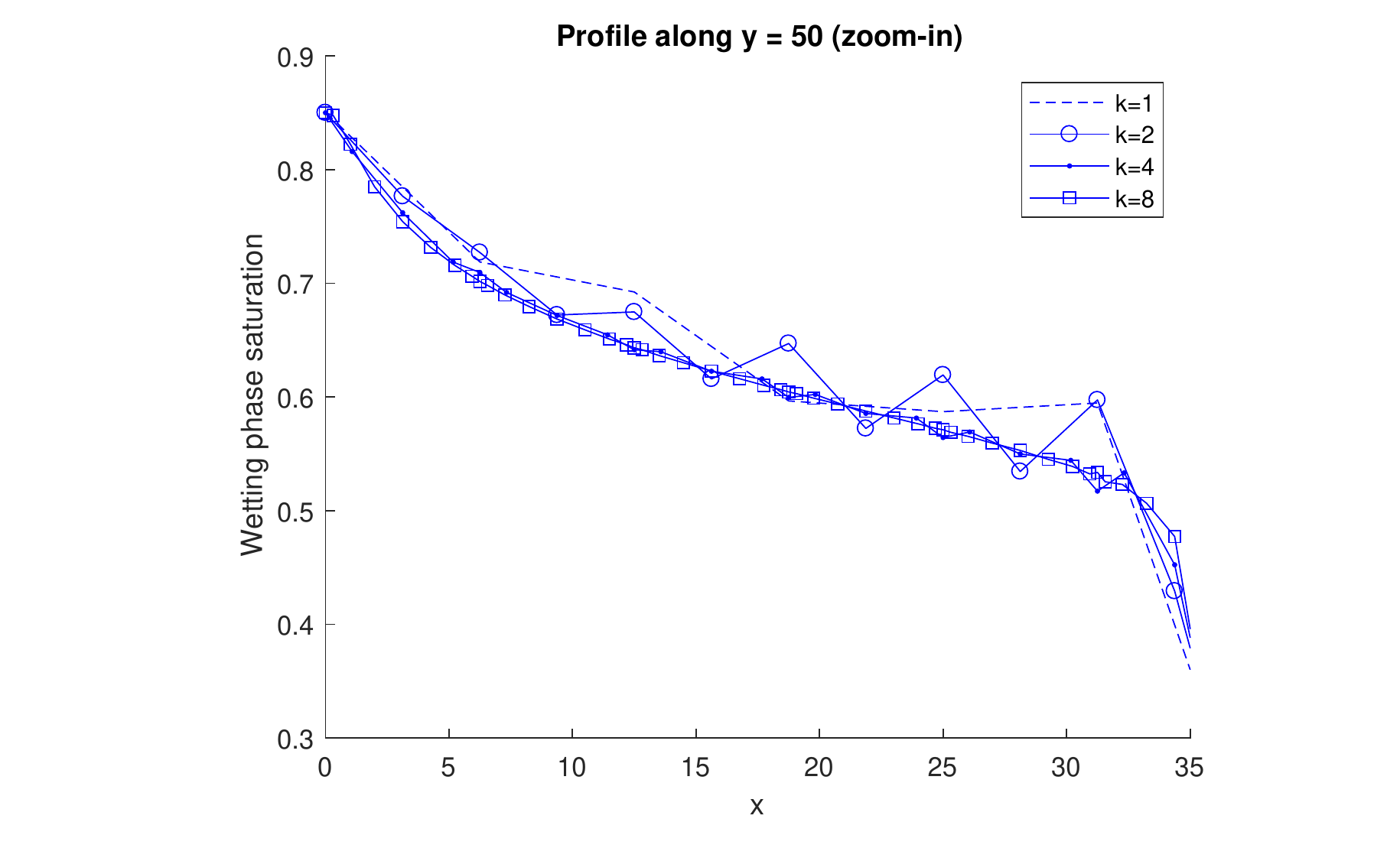} 
\end{minipage}
\begin{minipage}{0.45\textwidth}
\centering
\hspace*{-25ex}
\includegraphics[trim = 10mm 80mm 20mm 85mm, clip, scale = 0.65]{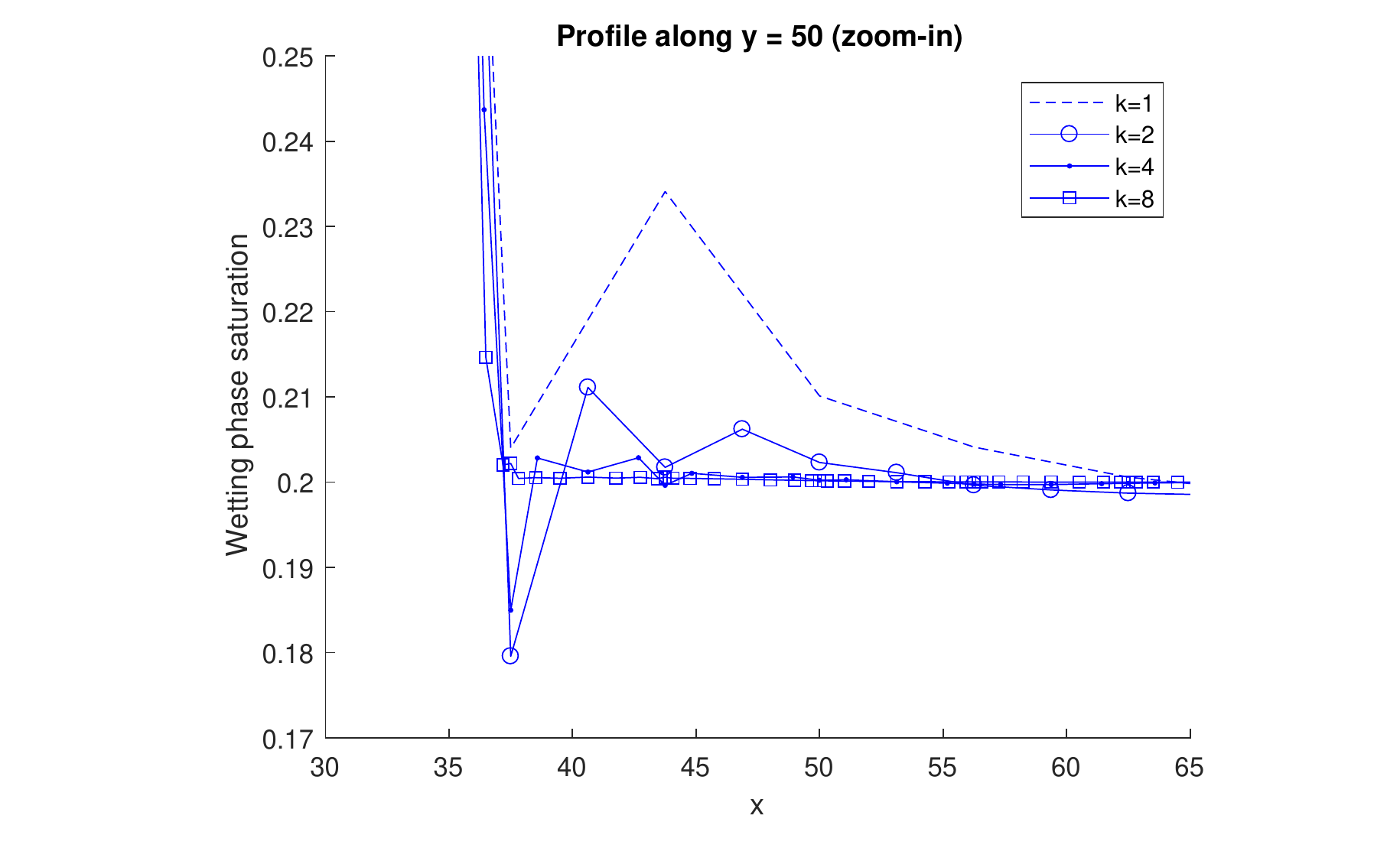}
\end{minipage}
\begin{minipage}{0.45\textwidth}
\centering
\hspace*{-15ex}
\includegraphics[trim = 10mm 80mm 20mm 85mm, clip, scale = 0.65]{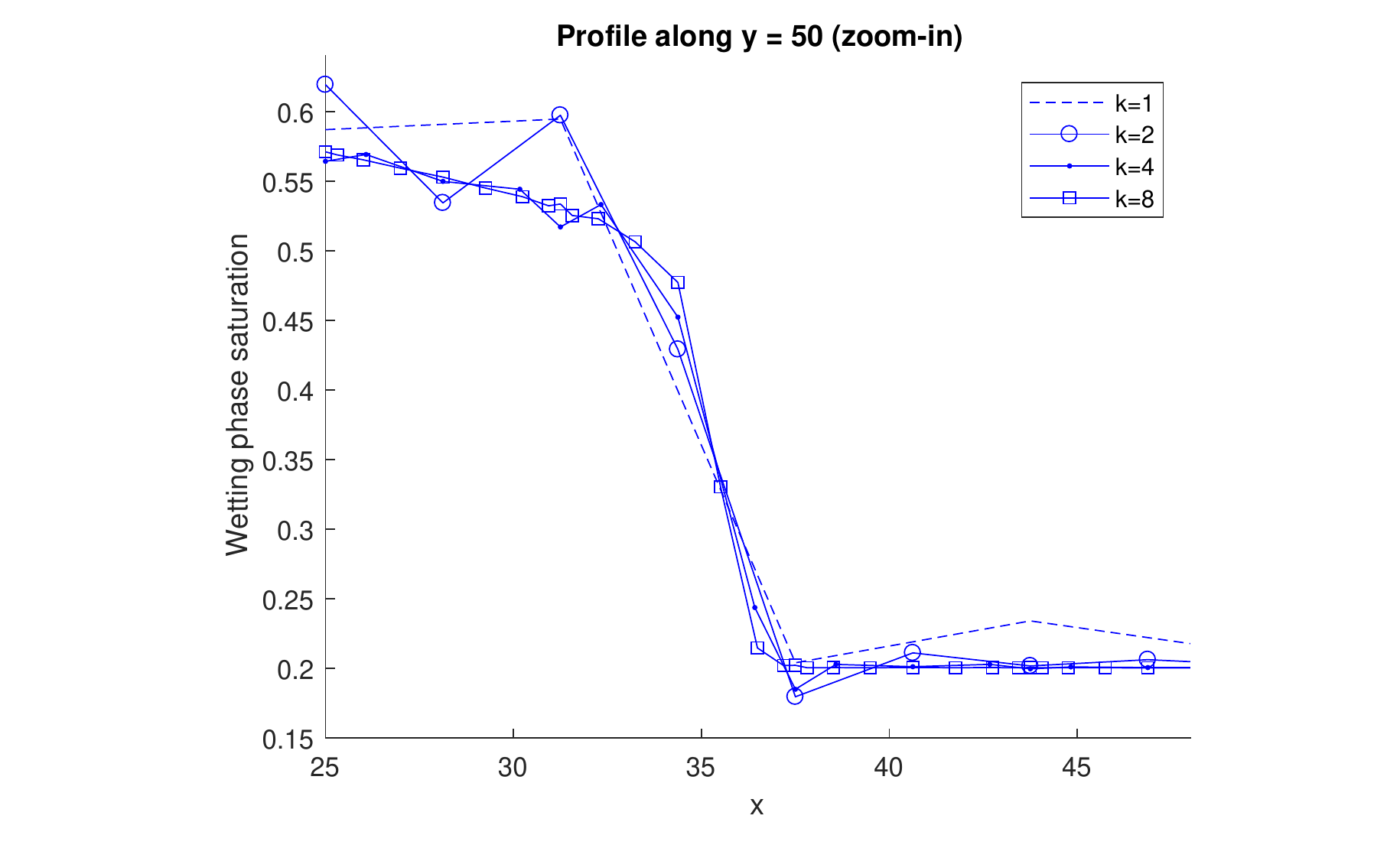} 
\end{minipage}
\caption{Inspection of the saturation profile along the line $y=50$.  The approximation converges as the polynomial order increases.}
\label{fig:utah_proflie_0}
\end{figure}

\begin{figure}[ht!]
\centering
\begin{minipage}{0.45\textwidth}
\centering
\hspace*{-25ex}
\includegraphics[trim = 10mm 80mm 20mm 85mm, clip, scale = 0.65]{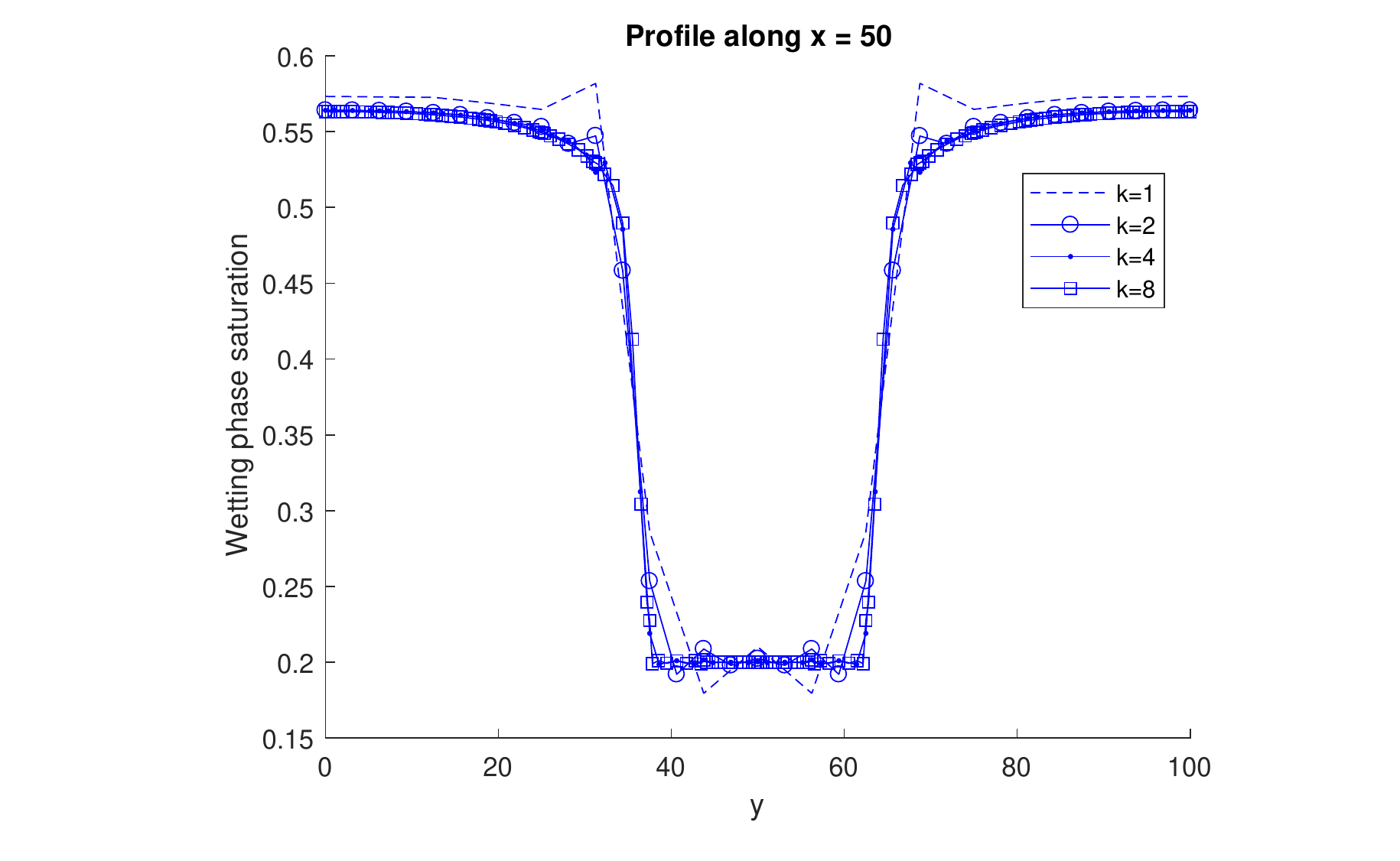}
\end{minipage}
\begin{minipage}{0.45\textwidth}
\centering
\hspace*{-15ex}
\includegraphics[trim = 10mm 80mm 20mm 85mm, clip, scale = 0.65]{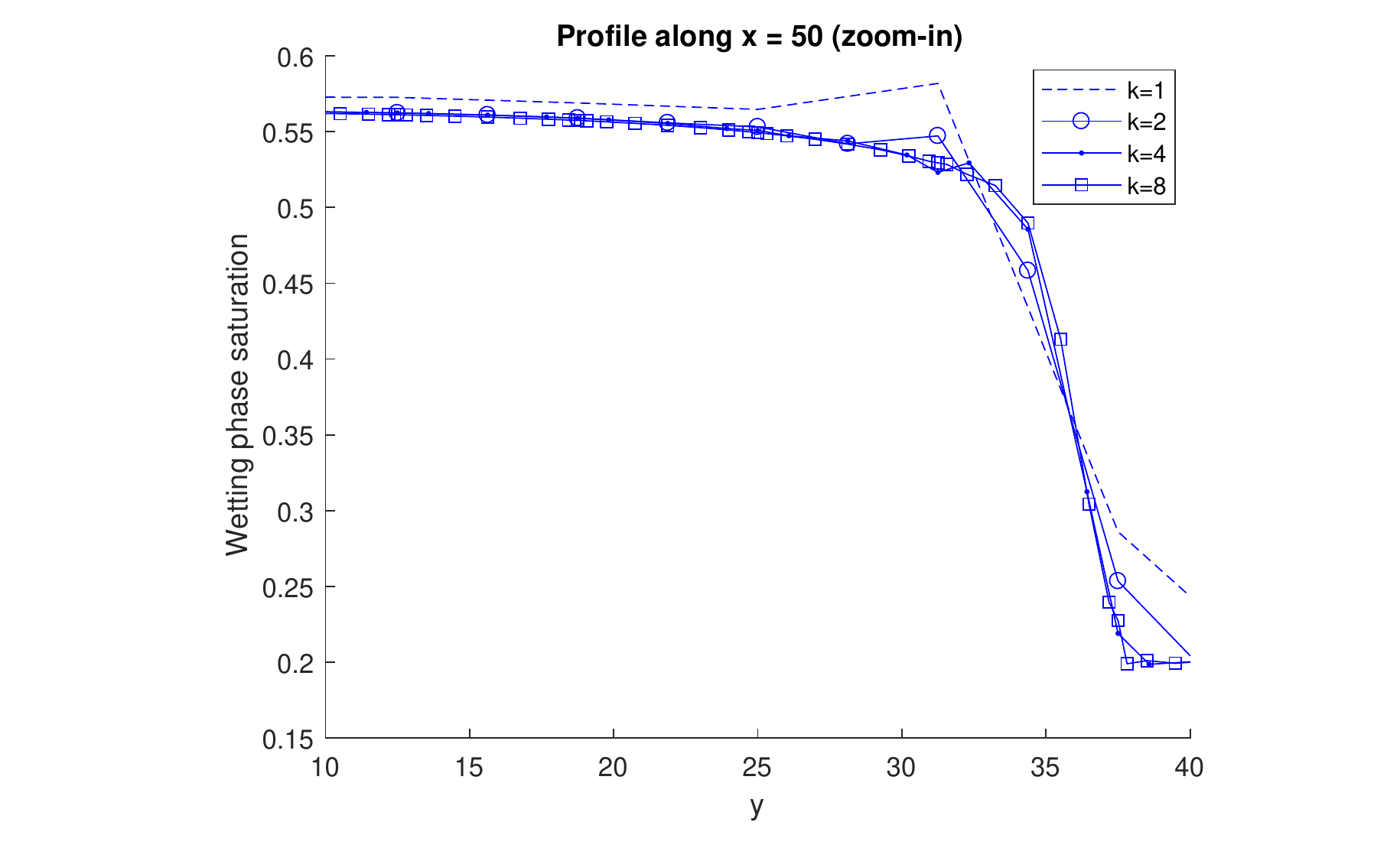} 
\end{minipage}
\begin{minipage}{0.45\textwidth}
\centering
\hspace*{-25ex}
\includegraphics[trim = 10mm 80mm 20mm 85mm, clip, scale = 0.65]{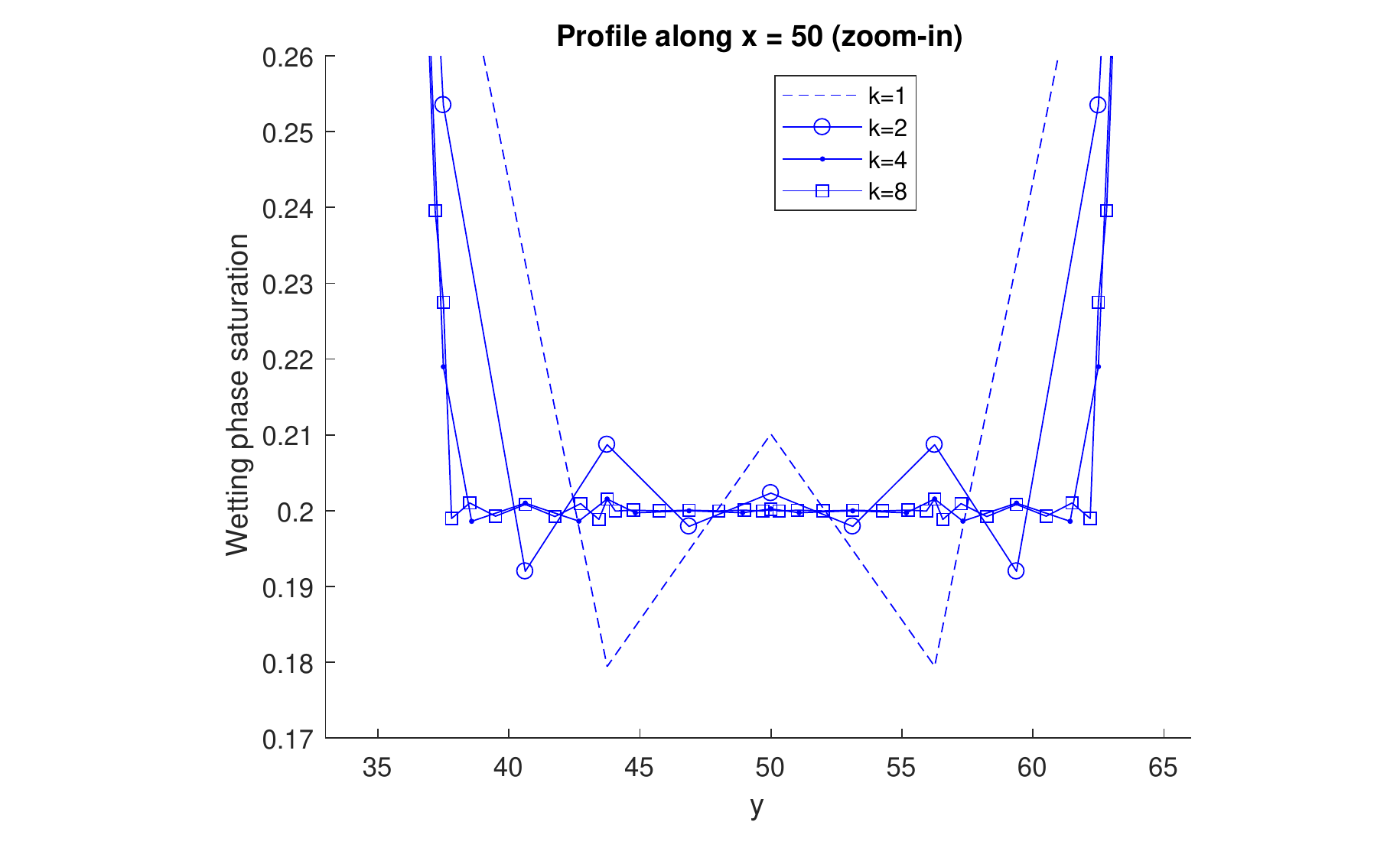}
\end{minipage}
\begin{minipage}{0.45\textwidth}
\centering
\hspace*{-15ex}
\includegraphics[trim = 10mm 80mm 20mm 85mm, clip, scale = 0.65]{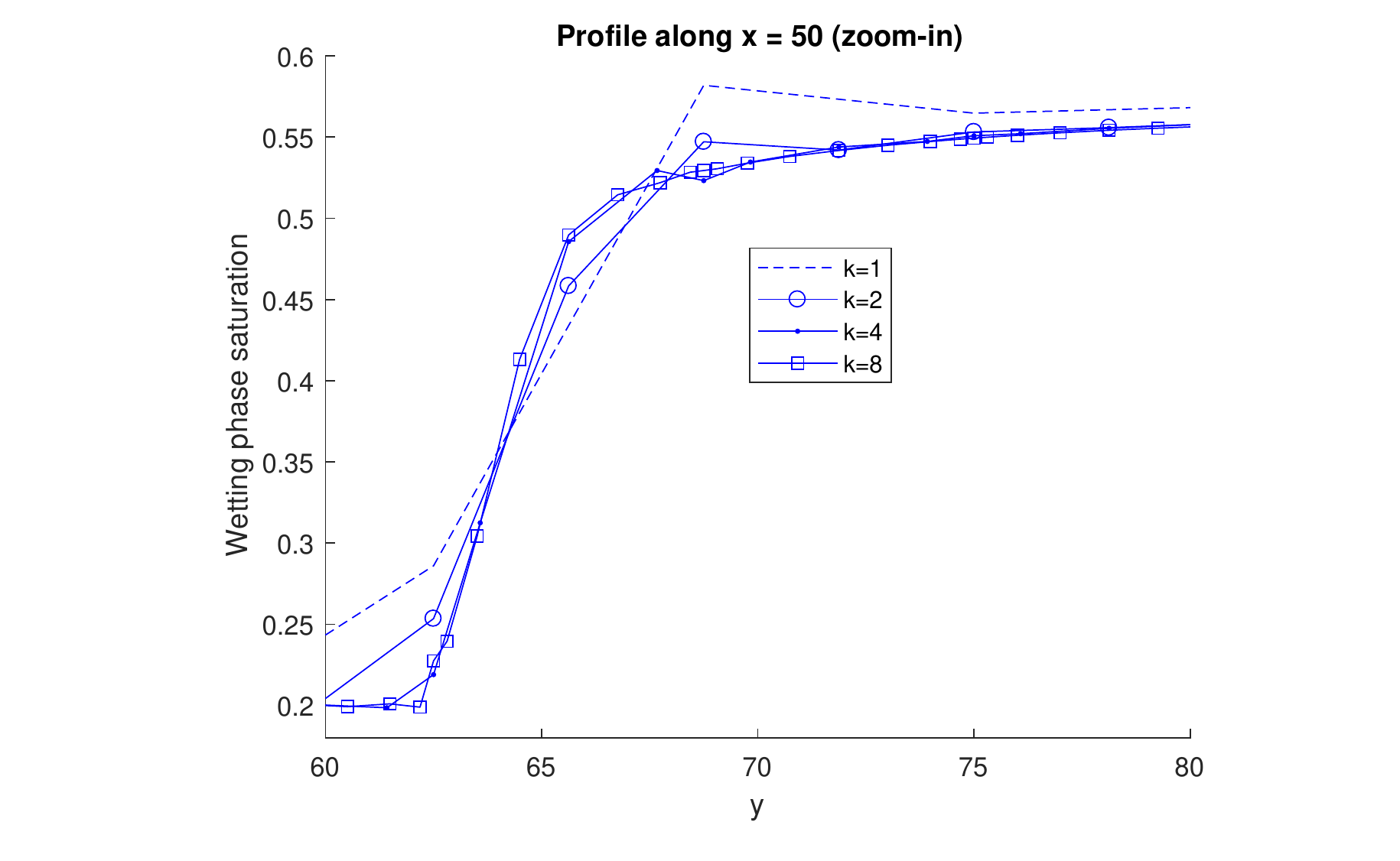} 
\end{minipage}
\caption{Inspection of the saturation profile along the line $x=50$.  The approximation converges as the polynomial order increases.}
\label{fig:utah_proflie_1}
\end{figure}

A heterogeneous test problem is explored here, with a similar set-up as in \cite{epshteyn2006solution}.  The permeability is
set as $K= 5\cdot 10^{-9}$ (m$^2$) in the domain $\Omega = [0,100]\times[0,100]$ (m), except in  the region  $[37.5,100]\times [37.5, 62.5]$, where $K = 5\cdot 10^{-13}$ (m$^2$).  Capillary pressure, relative permeabilities and viscosities are the same as in Section~\ref{sec:ex_Manu}.  At the boundary corresponding to $x=0$, we set  $p_{\mathrm{D}}=3\cdot 10^6$, and $s_{\mathrm{D}}=0.85$.  At the right boundary ($x=100$), we enforce $p_{\mathrm{D}}=1\cdot 10^6$ and $s_{\mathrm{D}}=0.2$.  The remaining boundaries are set as no flow.  A relatively coarse uniform mesh of 256 square elements is used.  The wetting phase floods the domain from the left boundary, and due to a pressure gradient flows from left to right.  As the wetting phase reaches the region of lower permeability, it is unable to invade, and must flow around it.
\\[12pt]
In Fig.~\ref{fig:utah_p_0} we plot the wetting phase saturation at 700 days for different polynomial orders.  The semi implicit method allows for large time steps, in our case we fix $\Delta t = 8$ days and use implicit Euler in time.  Overshoot and undershoot are visible for $k=1$ and $k=2$, and remain bounded.  As the polynomial order increases, overshoot and undershoot are no longer visible, and the saturation front is sharper.  Pressure contours are displayed in Fig.~\ref{fig:utah_p_0}.  The pressure contours do not appear to vary greatly by altering the polynomial order, however, the saturation curves are much more distinctive.
\clearpage
 \begin{figure}[ht!]
\centering
    \captionsetup{justification=centering}
   
    \subfigure[\small{$k=1$}]{
        \includegraphics[trim = 40mm 80mm 30mm 80mm, clip, scale = 0.5]{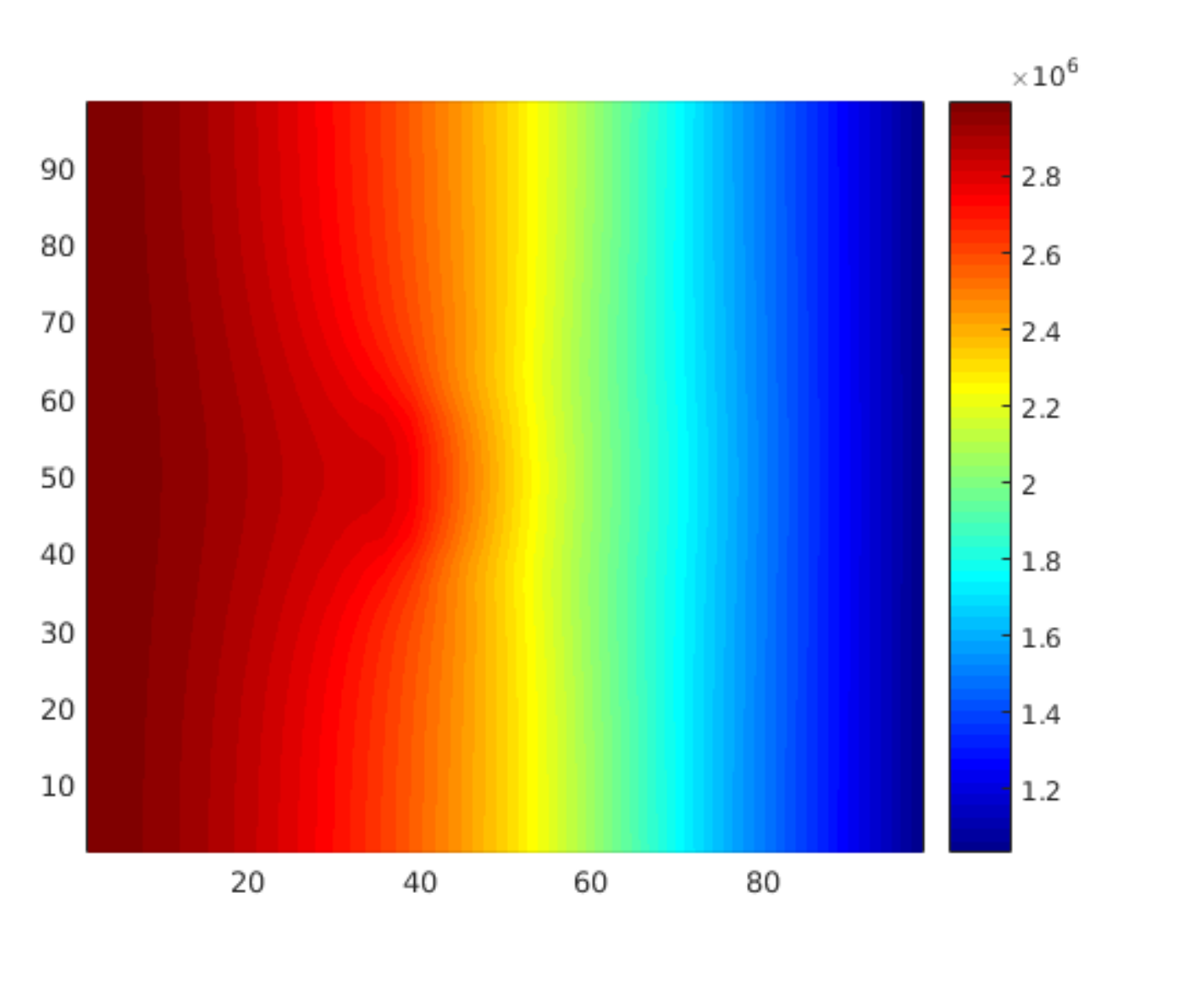}
    }
    \subfigure[\small{$k=2$}]{
        \includegraphics[trim = 40mm 80mm 30mm 80mm, clip, scale = 0.5]{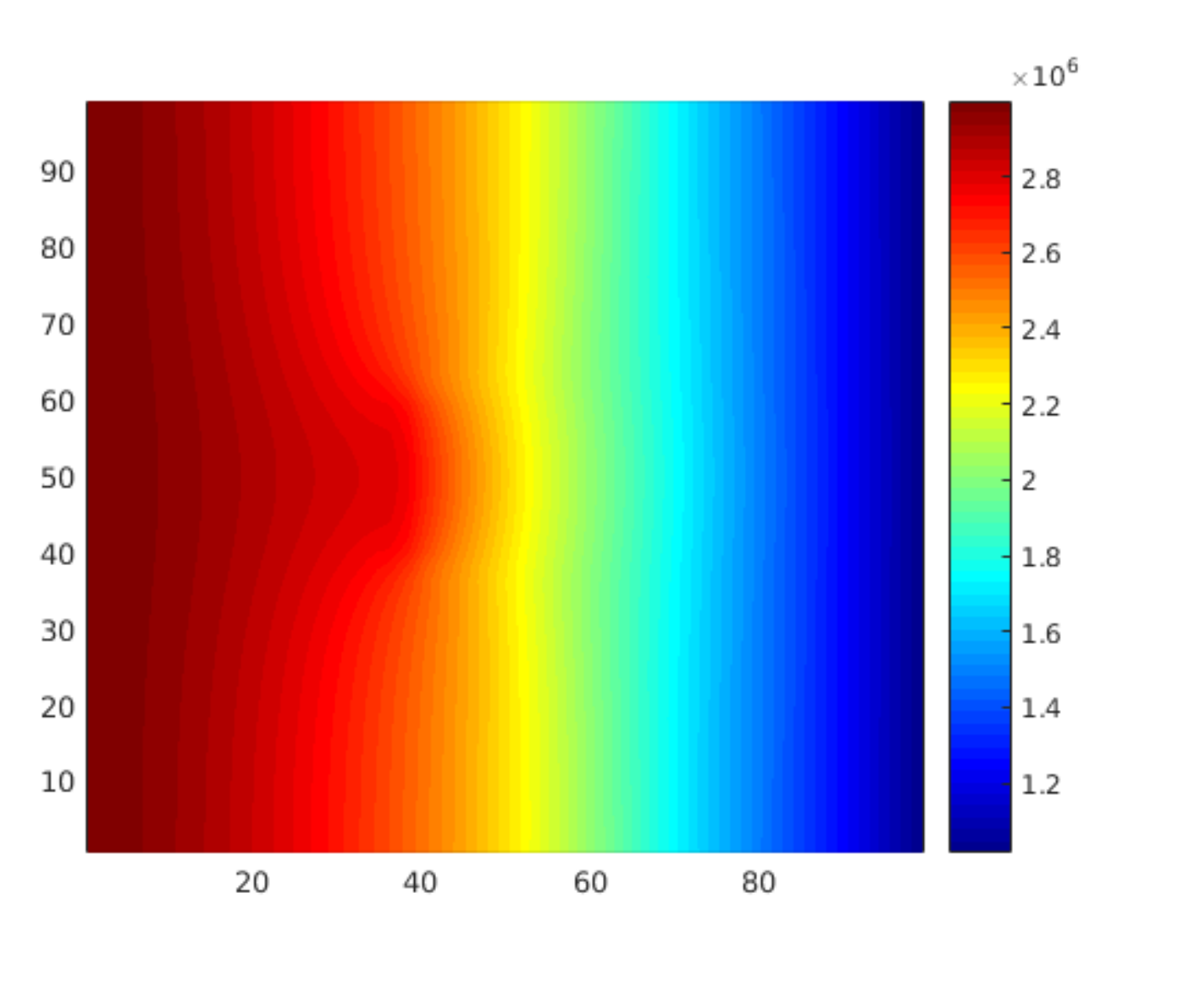}
    }    
    \\
    \subfigure[\small{$k=4$}]{
        \includegraphics[trim = 40mm 80mm 30mm 80mm, clip, scale = 0.5]{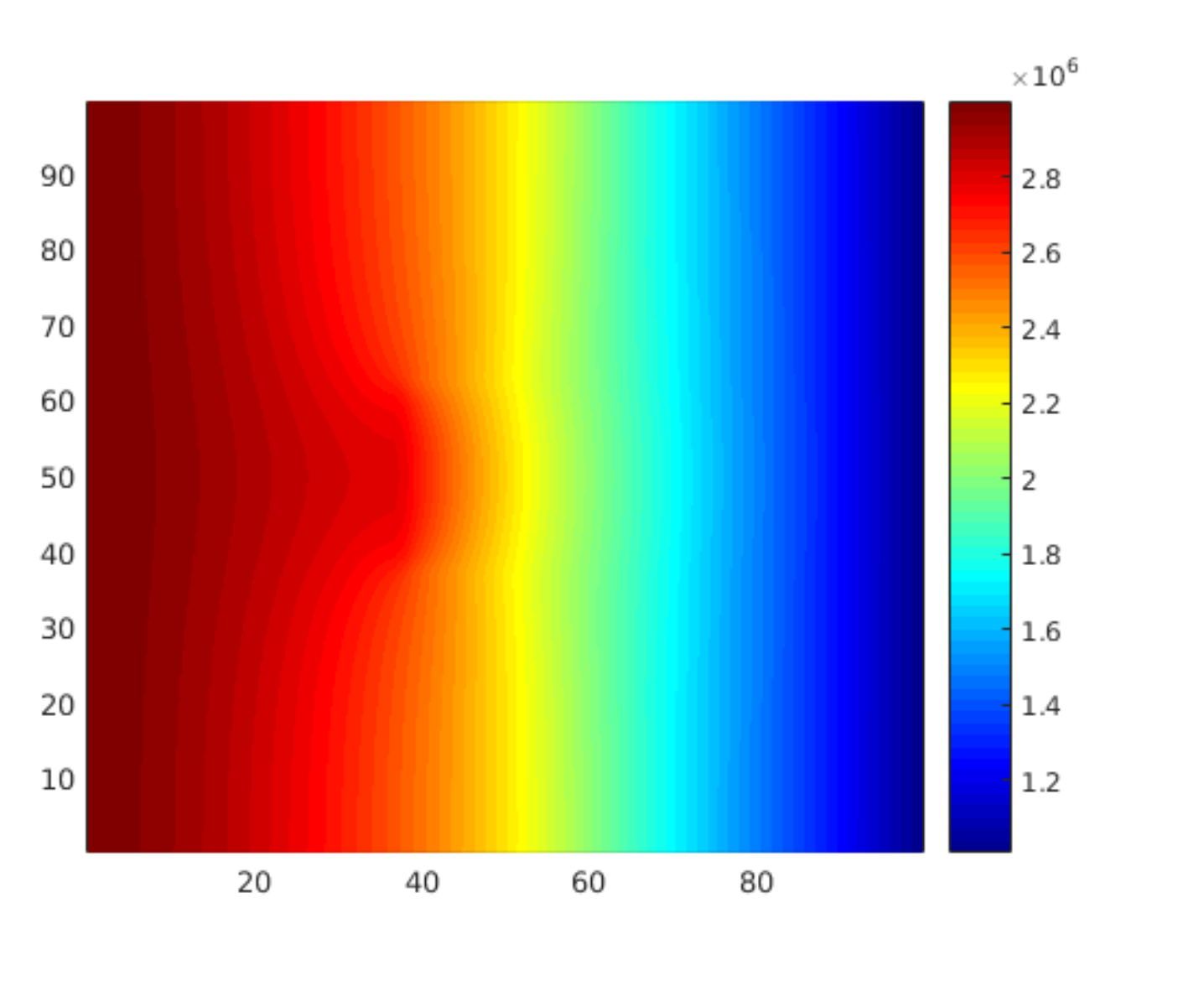}
    }
    \subfigure[\small{$k=8$}]{
        \includegraphics[trim = 40mm 80mm 30mm 80mm, clip, scale = 0.5]{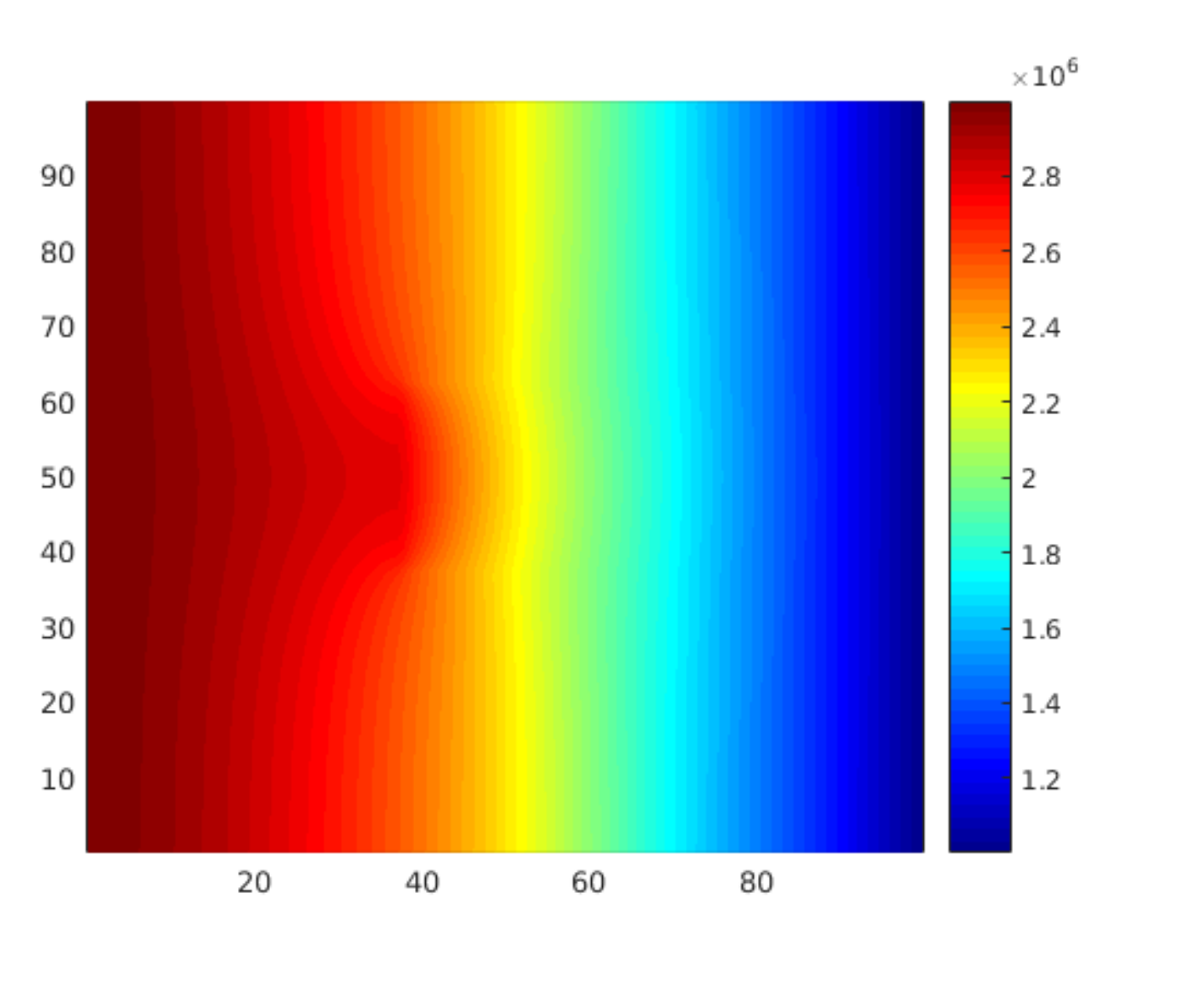}
    }      
\captionsetup{justification=justified}
\caption{Pressure contour plots at $T=700$ days for various polynomial orders.}
\label{fig:utah_pressure_0}
\end{figure}
\noindent  To examine the convergence of our method as we increase the polynomial degree, we plot the wetting phase saturation along the line $y=50$ in Fig.~\ref{fig:utah_proflie_0}.  Notice that this line intersects the region where the permeability drops by four orders of magnitude.  The spurious oscillations remain bounded and decrease with higher polynomial order.  Furthermore, as the polynomial order increases we see that the solution converges.
A similar plot for the vertical line $x=50$ can be found in Fig.~\ref{fig:utah_proflie_1}.  The conclusions remain the same in this case; higher order polynomials are beneficial.

\subsection{2D heterogeneous medium with realistic permeability}
\label{section_het2}
\begin{figure}[ht!]
\centering
    \captionsetup{justification=centering}

    \subfigure[\small{Wetting phase saturation at $t=150$ days}]{
        \includegraphics[trim = 40mm 80mm 30mm 90mm, clip, scale = 0.5]{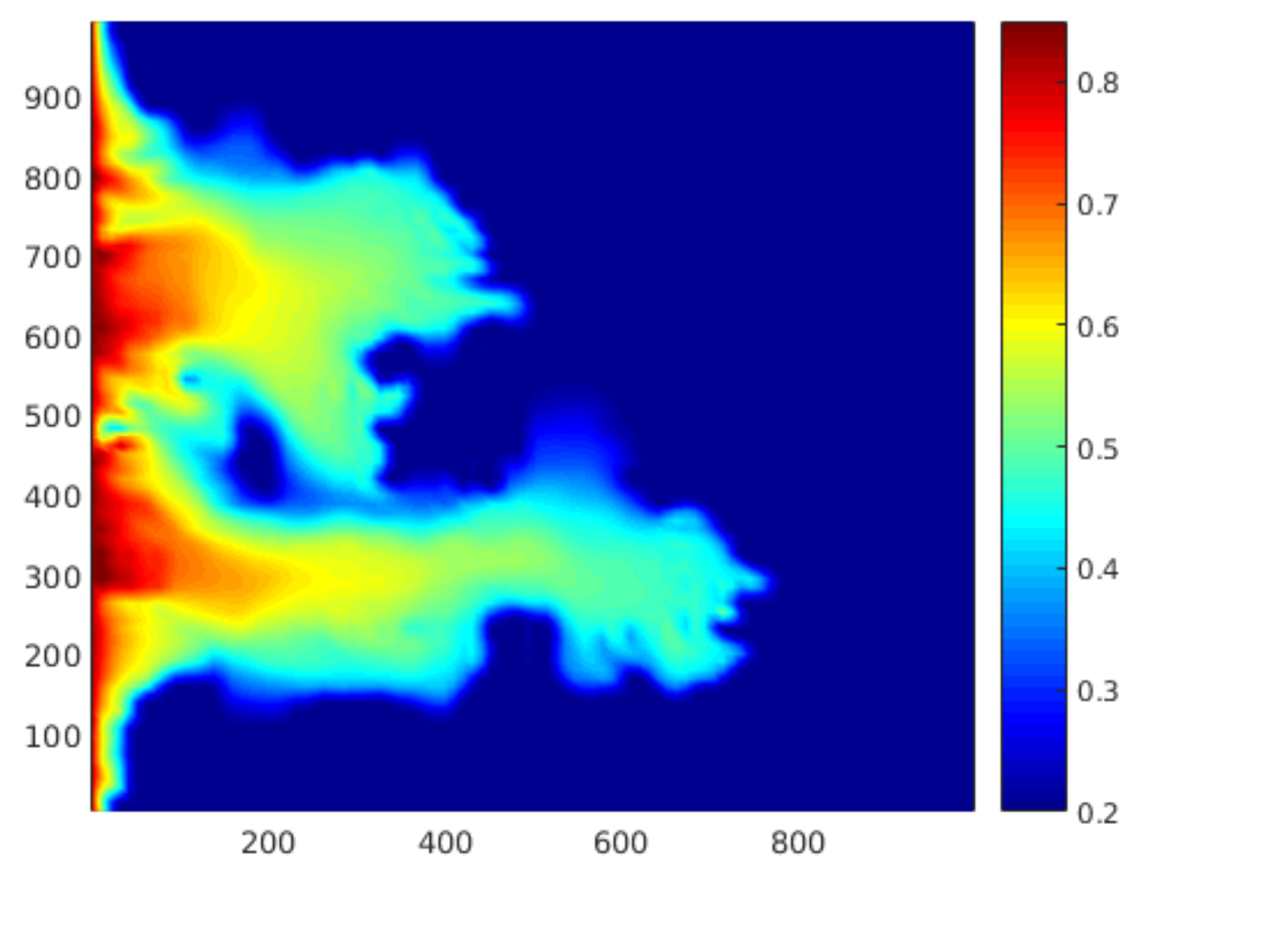}
    }
    \subfigure[\small{Layer 5 ($32\times 32$ grid, log scale)}]{
        \includegraphics[trim = 40mm 80mm 30mm 90mm, clip, scale = 0.5]{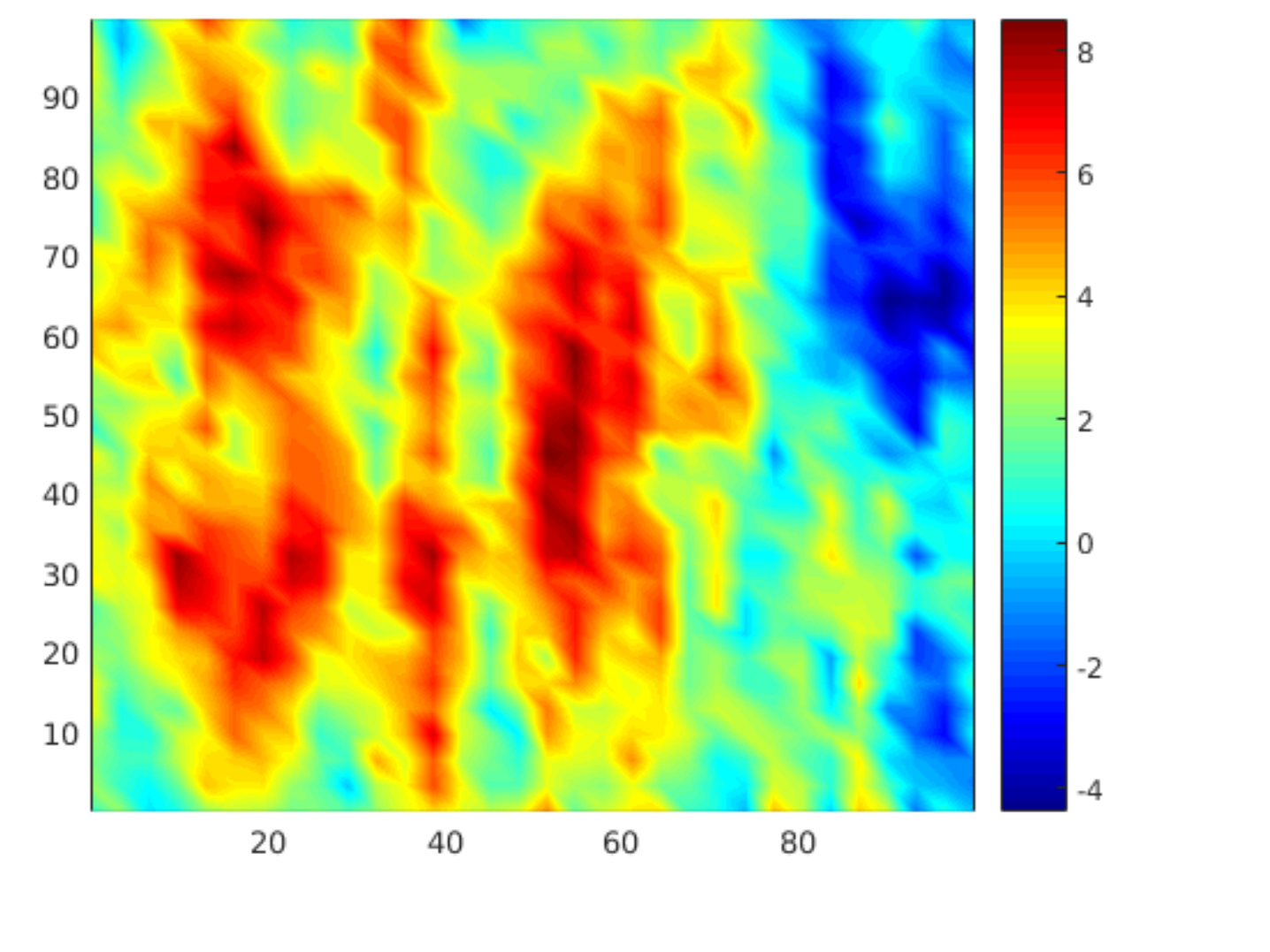}
    }    
    \\
    \subfigure[\small{Wetting phase saturation at $t=150$ days}]{
        \includegraphics[trim = 40mm 80mm 30mm 90mm, clip, scale = 0.5]{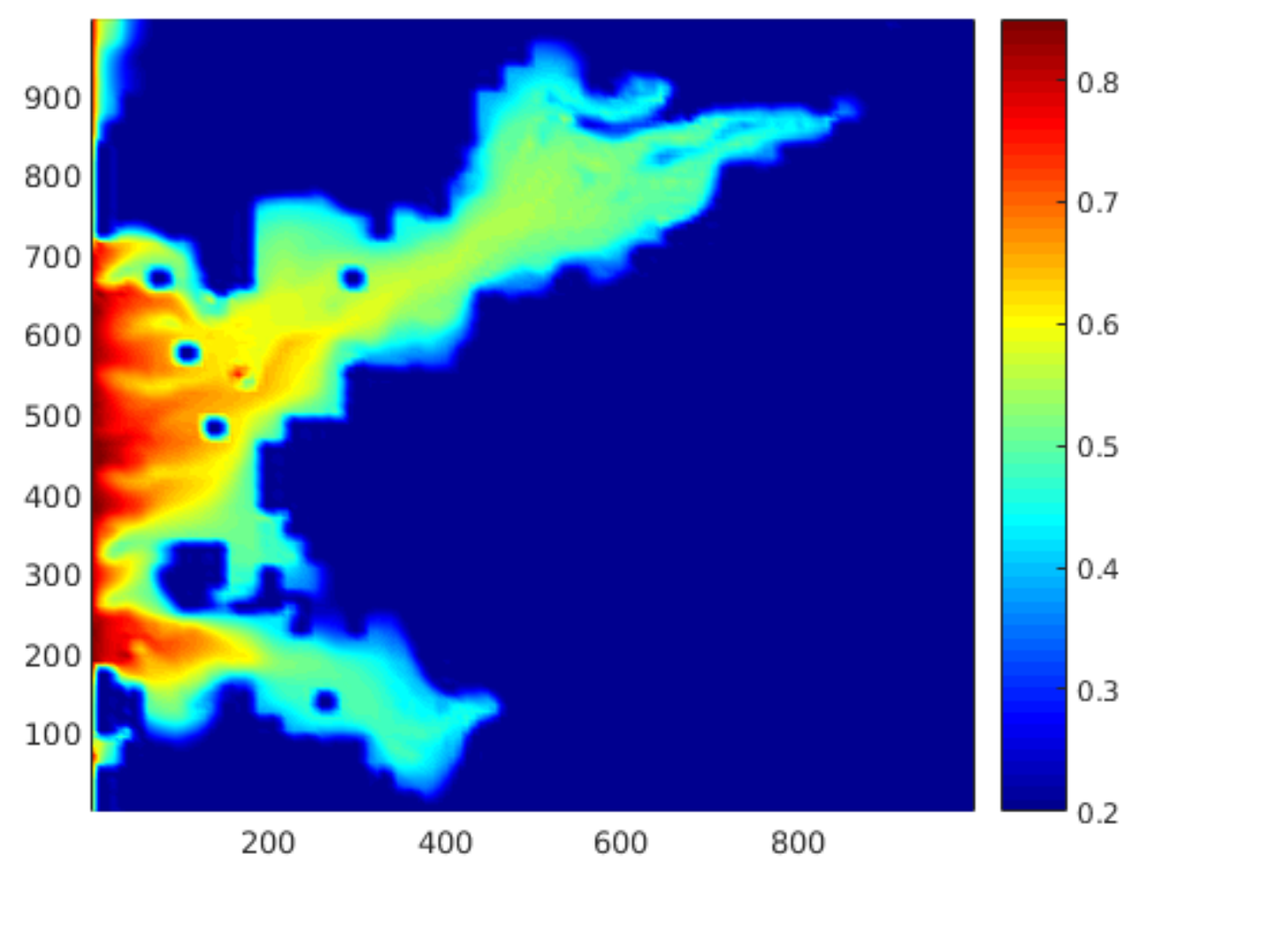}
    }
    \subfigure[\small{Layer 44 ($32\times 32$ grid, log scale)}]{
        \includegraphics[trim = 40mm 80mm 30mm 90mm, clip, scale = 0.5]{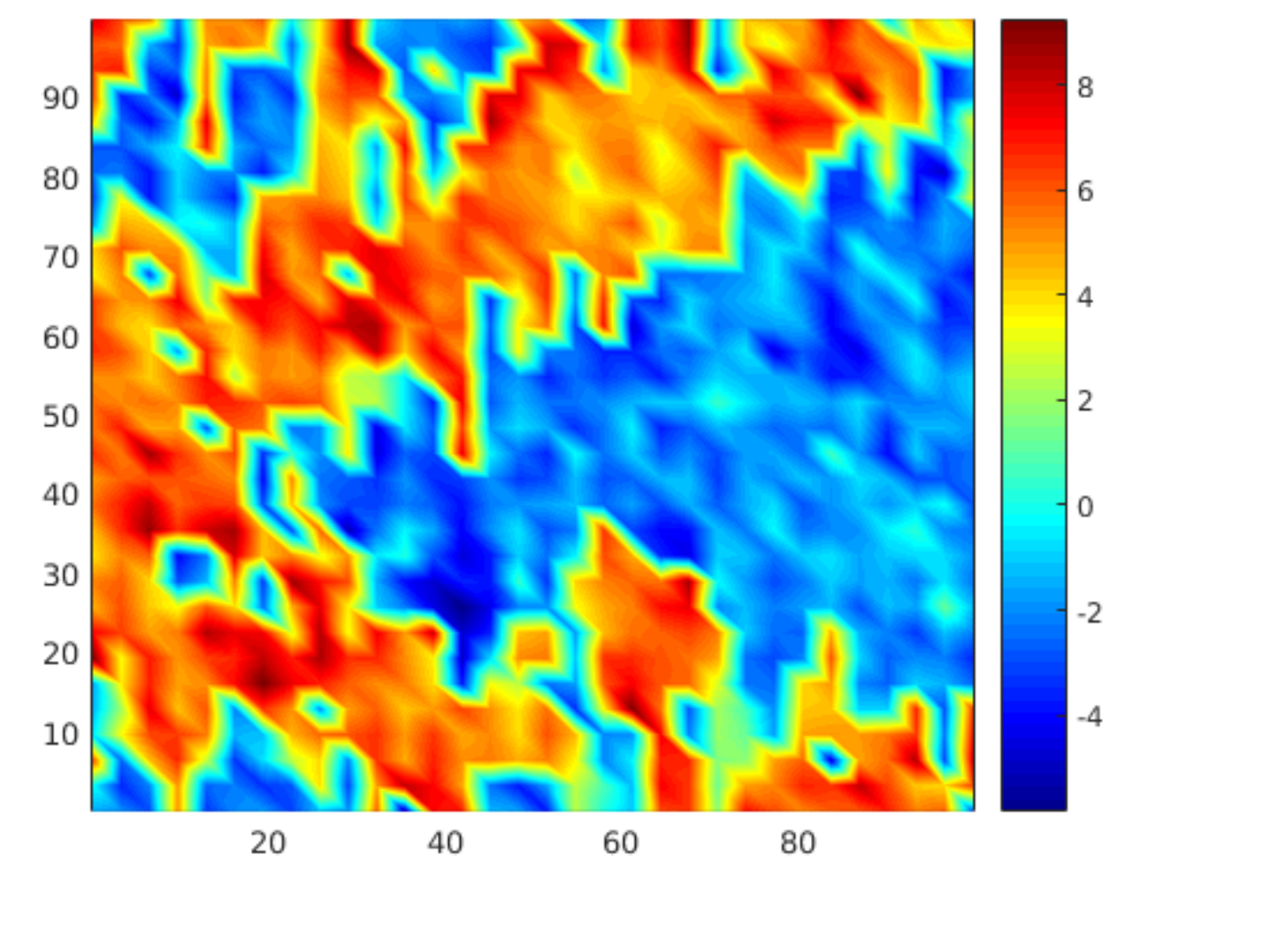}
    }    
    \\
    \subfigure[\small{Wetting phase saturation at $t=150$ days}]{
        \includegraphics[trim = 40mm 80mm 30mm 90mm, clip, scale = 0.5]{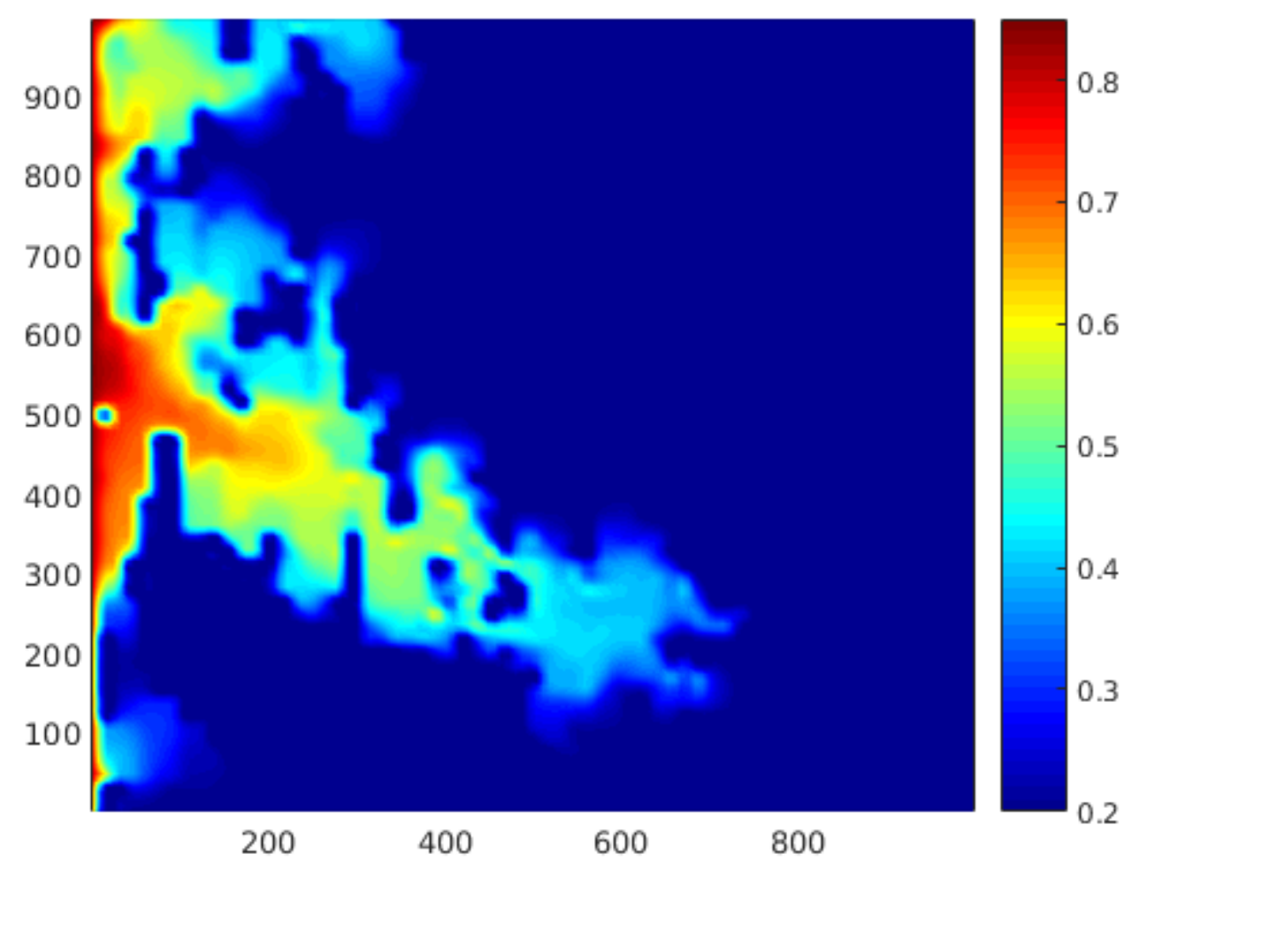}
    }
    \subfigure[\small{Layer 68 ($64\times 64$ grid, log scale)}]{
        \includegraphics[trim = 40mm 80mm 30mm 90mm, clip, scale = 0.5]{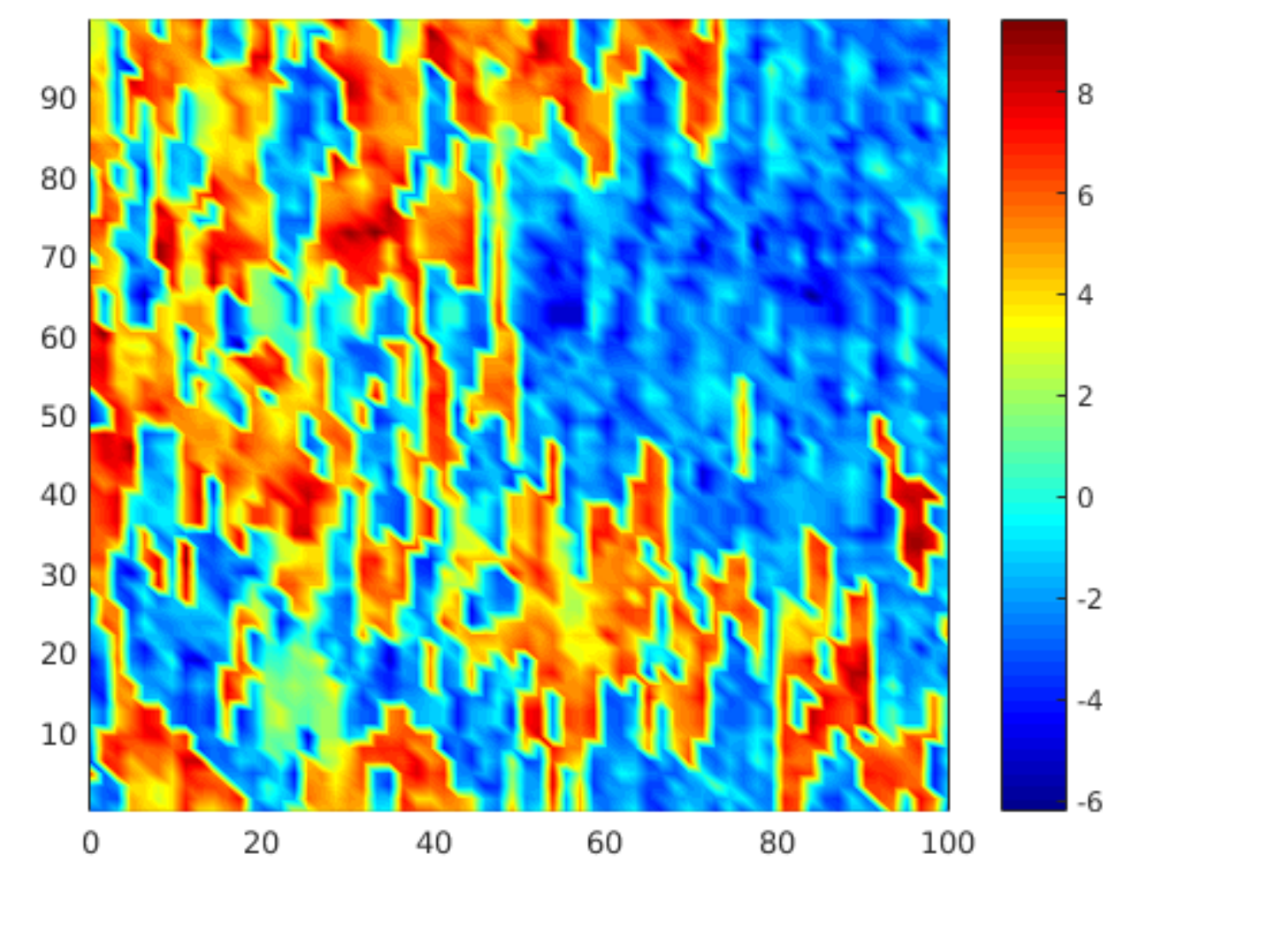}
    }      
\captionsetup{justification=justified}
\caption{Two-phase flow in highly heterogeneous media.   Uniform meshes with quadrilateral elements, and discontinuous piecewise quartic basis functions.}
\label{fig:Figure_spe_het}
\end{figure}

\noindent  All parameters are the same as in subsection~\ref{section_het1}, except for the domain that is larger ($\Omega=[0,1000]^2$) and for the permeability field.
For all simulations in this section we choose $k=4$.  The simulation is driven by boundary conditions, as described in the previous section.  Realistic permeability data is extracted from horizontal slices of the SPE10 CSP model 2 \cite{SPE10}.  We take $(32\times 32)$, $(32\times 32)$, and $(64\times 64)$ subsets of the data for vertical layers 5, 44, and 68, respectively.  Figs.~\ref{fig:Figure_spe_het} (b), (d), and (f) display the utilized highly heterogeneous permeability fields in log scale.  There are large channels in layer 44 whereas there are many small less-connected channels in layer 68.
\\[12pt]
Initially the reservoir has a wetting phase saturation of $0.2$.  Uniform meshes with quadrilateral elements are adopted: 1024 elements for Figs.~\ref{fig:Figure_spe_het} (a), (c) and 4096  elements for Fig.~\ref{fig:Figure_spe_het} (e).  We plot the wetting phase saturation after 150 days, which can be seen in Figs.~\ref{fig:Figure_spe_het} (a), (c), and (e).  It is evident that the wetting phase saturation avoids regions of low permeability which act as a barrier.  The permeability layers we selected range across the Tarbert  and Upper Ness formations, which demonstrates that the method is robust for permeabilities with different characteristics.

\section{Conclusions}
\label{sec:conclusions}
In this paper we presented a new method based on the hybridizable discontinuous Galerkin method for incompressible immiscible two-phase flow in porous media.  Numerical examples in 1D and 2D show that the method is high order accurate, and robust, even in the case of realistic discontinuous highly varying permeability.  Furthermore, we are able to take advantage of the HDG method, which is locally conservative, high order, and allows for a significant reduction in the total number of degrees of freedom through static condensation.  Static condensation enables us to consider polynomial orders that would be otherwise intractable for classical primal DG discretizations.  The method does not require penalization or slope limiters, and the stabilization factor does not depend on the polynomial order or mesh size.   

\section*{Acknowledgments}
This work was partially supported by NSF-DMS 1318348.  Fabien would like to acknowledge support from the Ken Kennedy Institute Graduate Fellowship Endowment, and Rice Graduate Education for Minorities (RGEM).

\newpage

  \newpage
 \clearpage
\newpage
\clearpage
\nocite{*}
\bibliographystyle{siamplain}
\bibliography{references}
\newpage
 \begin{table}[ht!]
\centering
\begin{tabular}{l  b  a  b  a  b  a  b}
\hline
 &    \multicolumn{3}{c}{$\| s_{wh}-s_w \|_{L^2(\Omega)}$} & \multicolumn{2}{c}{$\| s_{wh}^*-s_w\|_{L^2(\Omega)}$} 
&  \multicolumn{2}{c}{$\| {\bf q}_h -\nabla s_w\|_{L^2(\Omega)}$} 
\\ 
\cline{3-4} \cline{5-6} \cline{7-8}  
$k$  &$N$& Error & Rate & Error & Rate&  Error& Rate
\\ 
   1 &    4  &  1.05e-03 & -     & 6.84e-04 & -     & 7.28e-03 & - \\ 
     &    8  &  2.48e-04 & 2.089 & 1.12e-04 & 2.607 & 2.30e-03 & 1.661   \\
     &    16 &  6.08e-05 & 2.026 & 1.63e-05 & 2.777 & 6.61e-04 & 1.799   \\
     &   32  &  1.53e-05 & 1.992 & 2.22e-06 & 2.878 & 1.78e-04 & 1.889   \\
     &   64  &  3.85e-06 & 1.987 & 2.90e-07 & 2.936 & 4.65e-05 & 1.941  \\
     \\
   2&2   &3.94e-04&      -&   2.59e-04&       -&   2.51e-03&           -\\
    &4   &5.69e-05&  2.792&   2.28e-05&   3.503&   4.37e-04&   2.520\\
    &8   &7.39e-06&  2.945&   1.65e-06&   3.786&   6.36e-05&   2.781\\
    &16  &9.48e-07&  2.963&   1.12e-07&   3.876&   8.66e-06&   2.877\\
    &32  &1.20e-07&  2.981&   7.61e-09&   3.879&   1.17e-06&   2.887\\    
     \\
   3&2&  5.52e-05 &       -&  2.94e-05 &      - &  3.95e-04 &           -\\
   &4  & 3.35e-06 &  4.042 &  1.01e-06  & 4.863 &  2.81e-05  &  3.814\\
   &8 &  2.25e-07&   3.894 &  3.59e-08 &  4.817 &  2.00e-06  &  3.814\\
   &16&   1.47e-08&  3.934 &  1.20e-09  & 4.902 &  1.34e-07  &  3.900\\  
   &32&   9.48e-10&  3.954 &  4.014e-11 & 4.901 &  8.58e-09  &  3.963\\   
   \\
   4&2&  4.90e-06&       -&   1.42e-06 &      -&   2.80e-05 &    -\\
   &4 &  2.51e-07 &  4.284&   3.48e-08 &  5.350&   1.34e-06 &  4.380\\
   &8 &  9.58e-09 &  4.716&   6.09e-10 &  5.838&   4.87e-08 &  4.787\\
   &16&  3.20e-10&   4.904&   9.99e-12&   5.930&   1.62e-09&   4.905\\
   
   &32&  1.05e-11&    4.916&   1.69e-13&  5.882&   5.40e-11&   4.906\\
   \\
   5&2 &  7.49e-07&      -&   1.78e-07  &      -&   4.01e-06 &          -\\
   &4  & 1.98e-08 &  5.240&   1.56e-09  & 6.827 &  8.39e-08 &  5.581\\
   &8  & 3.20e-10 &  5.952&   1.15e-11  & 7.085 &  1.34e-09 &  5.959\\
   &16 &  5.19e-12&  5.946&   9.07e-14 &  6.991&   2.22e-11 &  5.924 \\ 
\hline
 \end{tabular}
\caption{Errors and convergence rates for the saturation, $s_{wh}$, the gradient, ${\bf q}_h$, and its post-processed approximation,  $s_{wh}^*$, on a Cartesian mesh of $N\times N$ elements.}
\label{tab1_sw}
\end{table}
 \begin{table}[ht!]
\centering
\begin{tabular}{b  b  a  b  a  b  b  b  a  b  a  b}
\hline
 &    \multicolumn{3}{c}{$\|p_{wh}-p_w\|_{L^2(\Omega)}$} 
&  \multicolumn{2}{c}{$\| {\bm u}_{th} - {\bm u}_t \|_{L^2(\Omega)}$} 
& \multicolumn{4}{c}{$\| p_{wh}-p_w\|_{L^2(\Omega)}$} 
&  \multicolumn{2}{c}{$\| {\bm u}_{th} - {\bm u}_t \|_{L^2(\Omega)}$}
\\
\cline{3-4} \cline{5-6} \cline{7-8}  \cline{9-10}  \cline{11-12}
$k$  &$N$& Error & Rate & Error & Rate& $k$ & $N$ & Error& Rate & Error& Rate
\\ 
   0& 2   &1.5217e-01   &    -    &3.2169e-03  &    -      &1&2  & 2.16e-02   &-      &2.04e-03   &-         \\
    & 4   &6.0507e-02   &1.330    &1.7914e-03  & 8.445e-01 &&4   & 7.49e-03   &1.529   &7.90e-04   &1.369\\
    & 8   &2.0184e-02   &1.583    &8.1952e-04  & 1.128     &&8   & 2.23e-03   &1.746   &2.39e-04   &1.720\\
    &16   &6.7061e-03   &1.589    &3.6293e-04  & 1.175     &&16  & 6.16e-04   &1.856   &6.75e-05   &1.827\\
    &32   &2.4355e-03   &1.461    &1.6977e-04  & 1.096     &&32  & 1.62e-04   &1.920   &1.83e-05   &1.881\\
    &64   &9.9662e-04   &1.289    &8.2987e-05  & 1.032     &&64  & 4.19e-05   &1.956   &4.84e-06   &1.919\\
   \\
   2&2  &3.90e-03  &-      &4.68e-04   &     -  &3&2  & 2.54e-04  &          -&   4.17e-05  &          -\\
   &4   &5.40e-04  &2.851   &6.77e-05   &2.791  &&4   &2.84e-05   &3.159  & 3.78e-06   &3.462\\
   &8   &7.34e-05  &2.880   &9.61e-06   &2.816  &&8   &2.56e-06   &3.474  & 3.16e-07   &3.581\\
   &16  &9.83e-06  &2.901   &1.35e-06   &2.826  &&16  & 1.82e-07   &3.814  & 2.28e-08  & 3.791\\
   &32  &1.28e-06  &2.936   &1.86e-07   &2.863  &&32   &1.19e-08  & 3.930  & 1.54e-09  & 3.885  \\    
   \\
   4&2   &7.88e-05   &  -    &8.99e-06   &    -   &5&2  &1.38e-05   &         -   &1.63e-06   &         -\\
   &4   &4.65e-06   &4.082   &5.20e-07   &4.112   &&4   &4.38e-07   &4.978   &5.51e-08   &4.888\\
   &8   &1.87e-07   &4.630   &2.17e-08   &4.578   &&8   &7.08e-09   &5.953   &9.59e-10   &5.843\\
   &16   &6.32e-09   &4.893   &7.60e-10   &4.840  &&16   &1.13e-10   &5.958   &1.64e-11   &5.871\\
   &32   &2.03e-10   &4.955   &2.52e-11   &4.911  &&32   &1.82e-12   &5.966   &2.73e-13   &5.905   \\
\hline
 \end{tabular}
\caption{Errors and convergence rates for $p_{wh}$ and ${\bf u}_{th}$, on a Cartesian mesh of $N\times N$ elements.}
\label{tab1_pw}
\end{table}
\end{document}